\documentclass[acmsmall]{acmart}

\usepackage{amsmath,amssymb,amsfonts}
\usepackage{algorithmic}
\usepackage{graphicx}
\usepackage{textcomp}
\usepackage{booktabs}
\usepackage{xcolor}
\usepackage{pifont}
\usepackage{tcolorbox}
\usepackage{multirow}
\usepackage{colortbl}
\usepackage{subfigure}
\usepackage{listings}
\usepackage{enumitem}

\usepackage{xurl}
\definecolor{color1}{RGB}{255,255,255}
\definecolor{color2}{RGB}{217,217,217}
\definecolor{myred}{RGB}{252, 142, 142}
\newcommand{\bench}{DesignBench}

\newcommand{\no}{\color{red!60!black}\ding{55}} 

\def\BibTeX{{\rm B\kern-.05em{\sc i\kern-.025em b}\kern-.08em
    T\kern-.1667em\lower.7ex\hbox{E}\kern-.125emX}}

\definecolor{bg}{HTML}{F8F9FB}  % FAFAFA
\definecolor{rowcolor}{HTML}{ECEFF4}
\definecolor{greenbg}{HTML}{D2E4DB}
\definecolor{added}{HTML}{C7E0D6}
\definecolor{removed}{HTML}{FBDBD8}

\AtBeginDocument{%
  \providecommand\BibTeX{{%
    Bib\TeX}}}

\setcopyright{acmlicensed}
\copyrightyear{2026}
\acmYear{2026}
\acmDOI{XXXXXXX.XXXXXXX}
\acmConference[Conference '26]{July,
  2026}{DC, USA}

\acmISBN{978-1-4503-XXXX-X/2018/06}

\begin{document}


% \title{\bench: A Comprehensive Benchmark for Front-end Code Generation}
% \title{\raisebox{-0.5ex}{\includegraphics[width=0.7cm]{figures/web-design.png}} \bench: A Comprehensive Benchmark for MLLM-based Front-end Code Generation}

\title{\bench: A Comprehensive Benchmark for MLLM-based Front-end Code Generation}

% \affiliation{%

\author{Jingyu Xiao}
\affiliation{%
  \institution{The Chinese University of Hong Kong}
  \city{Hong Kong}
  \country{China}
}
\email{jyxiao@link.cuhk.edu.hk}

\author{Ming Wang}
\affiliation{%
  \institution{The Chinese University of Hong Kong}
  \city{Hong Kong}
  \country{China}
}
\email{cd.melvin.wang@gmail.com}

\author{Man Ho LAM}
\affiliation{%
  \institution{The Chinese University of Hong Kong}
  \city{Hong Kong}
  \country{China}
}
\email{mhlam@link.cuhk.edu.hk}

\author{Yuxuan Wan}
\affiliation{%
  \institution{The Chinese University of Hong Kong}
  \city{Hong Kong}
  \country{China}
}
\email{yxwan9@cse.cuhk.edu.hk}

\author{Junliang Liu}
\affiliation{%
  \institution{The Chinese University of Hong Kong}
  \city{Hong Kong}
  \country{China}}
\email{Ljl0807123@gmail.com}

\author{Yintong Huo}
\authornote{Yintong Huo is the corresponding author.}
\affiliation{%
 \institution{Singapore Management University}
 \city{Singapore}
 \country{Singapore}}
\email{ythuo@smu.edu.sg}

\author{Michael R.Lyu}
\affiliation{%
  \institution{The Chinese University of Hong Kong}
  \city{Hong Kong}
  % \state{Texas}
  \country{China}}
\email{lyu@cse.cuhk.edu.hk}

% \author{\textbf{Project Page:} \url{https://webpai.github.io/DesignBench/}}

\renewcommand{\shortauthors}{Jingyu Xiao et al.}

\begin{abstract}
Multimodal Large Language Models (MLLMs) have demonstrated remarkable capabilities in automated front-end engineering, e.g., generating UI code from visual designs. However, existing front-end UI code generation benchmarks have the following limitations: (1) While framework-based development becomes predominant in modern front-end programming, current benchmarks fail to incorporate mainstream development frameworks. (2) Existing evaluations focus solely on the UI code generation task, whereas practical UI development involves several iterations, including refining editing, and repairing issues. 
(3) Current benchmarks employ unidimensional evaluation, lacking investigation into influencing factors like task difficulty, input context variations, and in-depth code-level analysis.

To bridge these gaps, we introduce \bench, a multi-framework, multi-task evaluation benchmark for assessing MLLMs' capabilities in automated front-end engineering. \bench~encompasses three widely-used UI frameworks (React, Vue, and Angular) alongside vanilla HTML/CSS, and evaluates on three essential front-end tasks (generation, edit, and repair) in real-world development workflows. \bench \ contains 900 webpage samples spanning over 11 topics, 9 edit types, and 6 issue categories, enabling detailed analysis of MLLM performance across multiple dimensions. Our systematic evaluation reveals critical insights into MLLMs' framework-specific limitations, task-related bottlenecks, and performance variations under different conditions, providing guidance for future research in automated front-end development. Our code and data are available at \url{https://github.com/WebPAI/DesignBench}.

\end{abstract}




% \keywords{Do, Not, Us, This, Code, Put, the, Correct, Terms, for,


\received{20 February 2007}
\received[revised]{12 March 2009}
\received[accepted]{5 June 2009}

\maketitle

\section{Introduction}

% Converting webpage design into functional UI code is a critical step for building websites, which can be labor-intensive and time-consuming. MLLMs have shown remarkable performance on visually rich code generation tasks \cite{yang2024swe}, which create new opportunities for the design-to-code task, i.e., automatically generating code from UI designs to replicate web page elements, layout, text, and colors. This widespread adoption of MLLMs has remarkably advanced practical development by not only automating repetitive tasks, but also improving the quality of code and accelerating the overall front-end development process. 

Converting webpage designs into functional UI code is a critical yet labor-intensive step in web development. MLLMs have demonstrated remarkable performance on visually rich code generation tasks \cite{yang2024swe, liu2025logomotion, Agile, li2024mmcode, chen2020wireframe, tang2025slidecoder}, creating new opportunities for automated design-to-code conversion that replicates webpage elements, layouts, text, and colors. 

% This advancement not only automates repetitive tasks but also improves code quality and accelerates front-end development.

% To understand the performance of MLLMs on front-end code generation, some evaluation benchmarks are proposed. 

Several benchmarks have been proposed to evaluate MLLMs on front-end code generation. These benchmarks typically involve either synthesizing webpage code via LLMs (e.g., WebSight~\cite{laurençon2024unlocking} and  Web2Code~\cite{yun2024web2code}), or curating code by cleaning real-world webpages (e.g., Webcode2M~\cite{gui2025webcode2m} and Design2Code~\cite{si2024design2code}).
%To assess MLLM's performance on front-end code generation, several evaluation benchmarks are proposed. For example, to create a high-quality dataset for the design-to-code problem, WebSight~\cite{laurençon2024unlocking} and Web2Code~\cite{yun2024web2code} leverage Large Language Models (LLMs) to synthesize webpage code, while Webcode2M~\cite{gui2025webcode2m} and Design2Code~\cite{si2024design2code} collect webpage code by cleaning real-world webpages. 
However, while these evaluation benchmarks are effective in measuring certain MLLM capabilities, they do not adequately represent the complex challenges faced by developers in real-world development scenarios.

(1) \textbf{Lack of front-end framework integration}. 
Current benchmarks fail to incorporate front-end frameworks, such as React, Vue, and Angular, which are integral to modern web development (as indicated in 2025 trends~\cite{web-framework}). Consequently, MLLMs' capabilities for practical framework-based website development remain unexplored.

% about 80\% of websites are developed using mainstream front-end frameworks (e.g., React, Vue). 

% Second, \textbf{insufficient tasks}. As shown in Fig.~\ref{fig:task}, the front-end development process encompasses three primary phases: initial front-end code \textit{generation} based on UI mockups, iterative code \textit{edit} to improve the UI design, and \textit{repair} the UI issues (e.g., compilation issue and display issue) in deployment environments. However, existing benchmarks only focus on the first phase, ignoring edit and repair.

\begin{figure}[t]
\centering
\includegraphics[width = .95\textwidth]{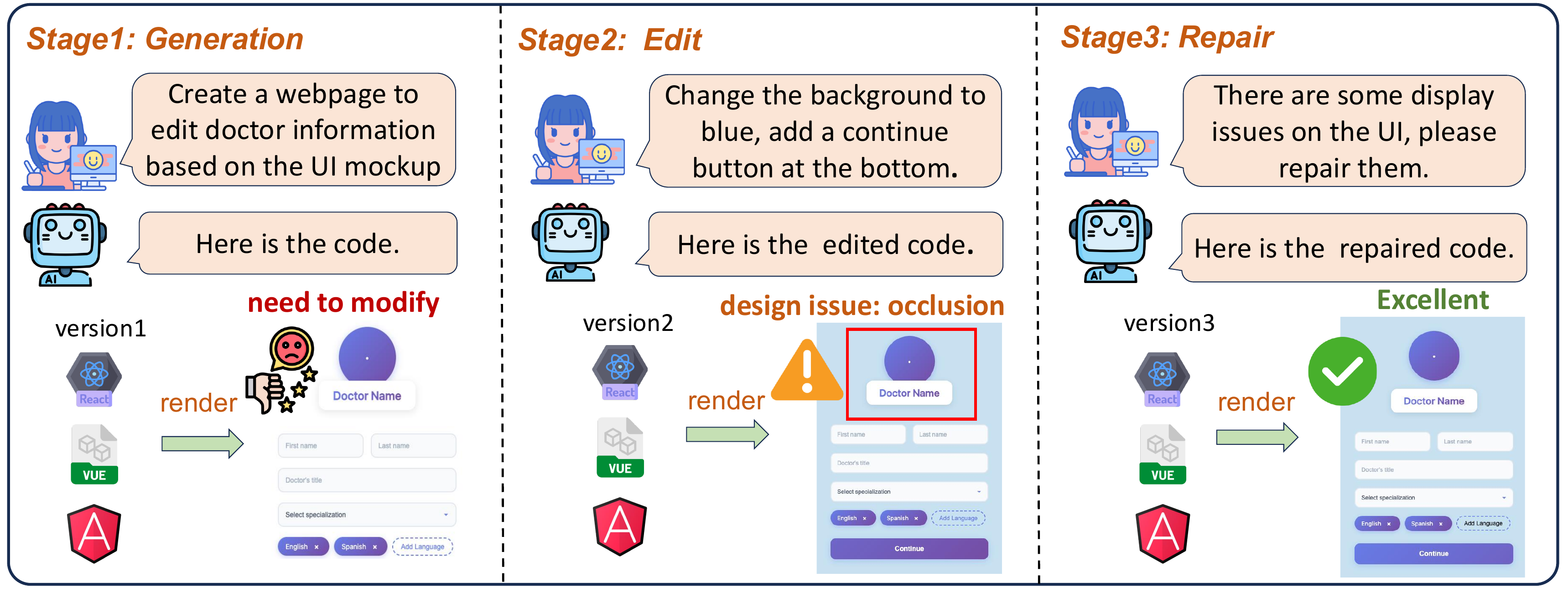}
\caption{Pipeline of MLLM-based automated front-end engineering.}
\label{fig:task}
\end{figure}

(2) \textbf{Insufficient task coverage}. Existing benchmarks inadequately cover the comprehensive spectrum of front-end development tasks. As illustrated in Fig.~\ref{fig:task}, development encompasses not only initial code generation from UI designs, but also iterative code editing for design refinement and repairing design-related issues~\cite{liu2020owl,liu2022nighthawk}. 
% three primary phases: initial code \textit{generation} from UI designs, iterative code \textit{edit} to refine designs, and \textit{repair} of design issues (e.g., display issues). 
However, current evaluations focus exclusively on the generation phase, neglecting the critical edit and repair phases.

(3) \textbf{Limited evaluation dimensions}. Most existing studies assess the overall quality of the generated webpage, lacking a detailed, multi-dimensional analysis of MLLM performance. Such an analysis could consider varying difficulty levels, input contexts, and code-level attributes like correctness and reusability. The comprehensive evaluation is crucial for identifying strengths and weaknesses of MLLMs, thereby fostering more reliable models for practical deployment.

% Existing benchmarks only focus on the 
% cover simple web design

To address these limitations, we propose \bench, a comprehensive multi-framework multi-task benchmark for evaluating MLLMs across multiple dimensions. \bench \ covers three popular front-end frameworks (React, Vue, Angular) and encompasses three key tasks: design generation, edit, and repair. The benchmark contains 900 webpage samples spanning more than 11 topics, 9 edit types, and 6 issue categories, enabling detailed analysis of MLLM performance across difficulty levels, context, and code dimensions.

% To address the above limitations, we propose \bench,  a multi-framework multi-task benchmark for evaluating the MLLMs from multi-dimensions. DesignBench involves three popular front-end frameworks (i.e., React, Vue and Angular) and contains three design tasks (i.e., design generation, edit and repair). \bench \ contains 900 webpage samples with more than 11 topics, 9 edit types and 6 issue types. We conduct a detailed analysis of MLLM in automating front-end engineering tasks from multiple dimensions, including difficulty, context, and code.

% Our extensive experiments yield several interesting findings. xxxxxxx

Our extensive experiments yield several key findings: 
(1) \textbf{Framework-specific limitations.} MLLMs exhibit substantially lower performance in framework-based development compared to vanilla HTML/CSS. They struggle with framework-specific syntax, such as JSX parsing (React), template syntax (Vue), and TypeScript architecture (Angular). Critically, these models fail to effectively leverage framework features and component-based paradigms, resulting in diminished code reusability. (2) \textbf{Task-specific bottlenecks.} Distinct performance bottlenecks were identified across the three core development tasks. Design generation task suffers from visual rendering inaccuracies and compilation errors, while design editing and repair tasks are primarily constrained by shortcomings in code localization and UI issue identification capabilities. (3) \textbf{Performance variability under different conditions}. MLLM efficacy significantly degrades under more challenging conditions, including large UI images for generation, complex instructions for editing, and severe UI issues for repair. Notably, an input context analysis on design edit and repair tasks reveals that code-only inputs consistently outperform image-only inputs. Multimodal combinations provide minimal additional improvement, suggesting that code representations convey more precise semantic information for these modification tasks than visual inputs.

% MLLMs are not good at framework-based development. MLLMs exhibit fundamental deficiencies in framework-specific syntax comprehension, including JSX parsing difficulties in React, template syntax challenges in Vue, and TypeScript component architecture struggles in Angular. MLLMs are not good at leveraging framework featured syntax and component-based implementation produce reusable front-end code.

% MLLMs demonstrate significant framework-specific limitations, excelling with vanilla HTML/CSS but struggling considerably with modern framework implementations, particularly Angular's TypeScript architecture, while facing persistent challenges in JSX parsing for React and template syntax for Vue, ultimately revealing fundamental deficiencies in component-based implementation and framework-specific syntax comprehension that lead to frequent compilation errors. 

\begin{table*}[ht]
\centering
\setlength{\tabcolsep}{0.2em}
\caption{Comparison of existing UI-to-Code benchmark and \bench. Vanilla refers to plain HTML/CSS.}
\label{tab:bench_compare}
\resizebox{\textwidth}{!}{
\begin{tabular}{llllll}
\toprule
Benchmark   & Sample & Source & Framework & Task & Dimension \\
\midrule
Pix2code \cite{beltramelli2018pix2code}    & 1742    & Synthetic        & Vanilla       & Generation  & \no       \\
WebSight \cite{laurençon2024unlocking}    & 823K    & Synthetic        & Vanilla       & Generation  & \no       \\
Web2Code \cite{gui2024vision2ui}   & 60K    & Synthetic        & Vanilla      & Generation  & \no       \\
WebCode2M \cite{gui2025webcode2m}   & 20K    & Real-world        & Vanilla      & Generation  & \no      \\
Design2Code \cite{si2024design2code} & 484    & Real-world        & Vanilla       & Generation  &  \no      \\
\midrule
\bench & 900    & Real-world        & Vanilla/React/Vue/Angular       & Generation/Edit/Repair  & Difficulty/Context/Code/Failure Type    \\
\bottomrule
\end{tabular}}
\end{table*}

% visual complexity, instruction difficulty and code  

% When solving competitive programming problems, programmers usually write an initial program, execute some example test cases, and refine the code based on the test results. 
% In this process, a programmer can take key information (e.g, program outputs or compile/runtime error message) from the test results, which helps them debug the program. We instantiate this idea by adopting a similar pipeline with a neural-based editor (in Figure 1(a))

Our main contributions are summarized as follows:

% \begin{itemize}
%     \item We present the first multi-framework multi-task benchmark for evaluating the MLLM's capabilities on automated front-end engineering.

%     \item We comprehensively evaluate the performance of nine MLLMs on three tasks from three dimensions, namely, difficulty (i.e., the visual complexity, instruction complexity and issue severity); context (i.e., code and image); and code (i.e., code correctness and reusability).

%     \item Our findings reveal MLLMs' framework-specific limitations, bottlenecks on three tasks and performance under different conditions.
% \end{itemize}

\begin{itemize}[leftmargin=*]
\item We introduce the first multi-framework, multi-task benchmark for evaluating MLLMs' capabilities in automated front-end engineering across HTML/CSS, React, Vue, and Angular frameworks.
\item We conduct an extensive evaluation of nine leading MLLMs across three fundamental tasks. The analysis considers multiple dimensions, including task difficulty (visual complexity, instruction complexity, and issue severity), input context modalities (code-only, image-only, and multimodal), and code metrics (correctness and reusability).

\item We reveal key insights into MLLMs' framework-specific limitations, identify task-dependent performance bottlenecks, characterize performance variations under diverse conditions and identify 22 types of failure of the three tasks. These results provide important guidance for future research and practice in front-end development.

% \item To facilitate further research, we publicly release the code and data on Github~\cite{DesignBench}.
\end{itemize}

% \begin{figure*}[ht]
% \centering
% \includegraphics[width = .9\textwidth]{figures/Benchmark.pdf}
% \caption{The task.}
% \label{fig:overview}
% \end{figure*}

% After collecting API data, we categorize the API suggestion scenarios into three scenarios including
% “when to use”, “which to use”, and “how to use” as shown in Figure
% 1. These scenarios simulate different situations where developers
% use APIs in programming practice and evaluate the capabilities of
% current LCMs to generate accurate suggestions in these scenarios
% We categorize the automa

\section{Background}

\subsection{Web Development Process}

Web application front-end development process includes the following stages:
(1) \textbf{Design stage}: designers create high-fidelity mock-ups using prototyping tools such as Sketch~\cite{Sketch} and Axure~\cite{Axure} during design stage. (2) \textbf{Development stage}: it involves transforming the design concepts into a functional application through coding. The development stage typically consists of the implementation of the GUI and the underlying functionalities. As shown in Fig.~\ref{fig:task}, the developer first prompts the MLLM to generate the front-end code based on the UI-Mockup. However, the version one code did not fully comply with the UI-Mockup. Then the user asked MLLM to edit the code to change the background color and add buttons to generate version two. Then during the actual deployment process, the front-end engineer found a display issue on the UI, that is, a collision between the avatar and the name, and asked MLLM to fix it and generate version three code.

\subsection{Front-end Web Development Framework}

Front-end frameworks are collections of pre-written, reusable code that provide a foundation for building the user interface of a website. These frameworks can save time and improve the quality of the final product by providing a standardized set of components and templates for front-end development. In 2025, React, Vue, and Angular have emerged as the predominant frameworks in web development~\cite{web-framework}, with 39.5\%, 15.4\%, and 17.1\% popularity~\cite{FrontendFrameworksPopularity}, respectively.

\section{Related Work}

\subsection{UI Code Generation}
UI code generation techniques fall into three categories: Deep Learning (DL) based methods, Computer Vision (CV) based methods, and Multimodal Large Language Model (MLLM) based methods. (1) DL-based methods: \cite{acsirouglu2019automatic, cizotto2023web, moran2018machine, Xu2021Image2e, Chen2018FromUI} leverages CNNs to automatically prototype software GUIs. Pix2code \cite{beltramelli2018pix2code} utilizes CNNs and LSTM to extract features from GUI images to generate a domain-specific language (DSL). \cite{chen2022code} implements an encoder-decoder framework with an attention mechanism to generate the DSL. (2) CV-based methods: Sketch2Code \cite{jain2019sketch2code} inputs hand-drawn sketches into object detection models to learn the object representation, which is read by the UI parser to generate code for targeted platforms. REMAUI \cite{nguyen2015reverse} identifies user interface elements via optical character recognition (OCR) techniques and then infers a suitable user interface hierarchy and exports the results as source code. (3) MLLM-based methods~\cite{lu2025misty,wan2024automatically,zhou2024bridging,designrepair, wan2025automatically, xiao2025efficientuicoder, xiao2026comuicoder, dang2025envisioning}: to solve the element omission distortion and misarrangement problems during UI code generation, DCGen \cite{wan2024automatically} proposes a divide-and-conquer-based approach to generate the code of the submodules separately and then assemble them to construct the full webpage based on MLLMs. DeclarUI \cite{zhou2024bridging} applies the element segmentation method to accurately generate elements and page transition graphs to prompt MLLMs to generate mobile app UI with jump logic. EfficientUICoder~\cite{xiao2025efficientuicoder} accelerates UI code generation by input and output token compression. ComUICoder~\cite{xiao2026comuicoder} apply semantic-aware segmentation and merging methods for reusable UI code generation. \textbf{While prior works achieve decent performance in UI-to-code, none of them address the conventional framework-based development and other essential front-end procedures like design edit and repair.}

\subsection{Benchmarks for UI Code Generation}

Many benchmarks~\cite{gu2024cruxeval, lam2025codecrash, jain2024livecodebench, zhuo2024bigcodebench} have been proposed to evaluate the code generation, understanding and reasoning capabilities of LLMs, but there are only a limited number of benchmarks for UI code. To improve MLLMs' UI-to-Code capabilities, Pix2code introduced a domain-specific language (DSL) and leveraged CNN and LSTM architectures to translate UI mockups into DSL code. However, due to the inherent limitations of DSL approaches, Pix2code suffers from poor scalability and limited real-world applicability. WebSight advanced the field by synthesizing high-quality HTML code for training, while Web2Code~\cite{yun2024web2code} proposed the Webpage Code Generation Benchmark (WCGB) to systematically evaluate MLLMs' HTML parsing capabilities. Despite these contributions, both benchmarks rely on synthetic data, which may not capture the complexity and variability of real-world web development scenarios. Design2Code~\cite{si2024design2code} addressed this limitation by manually curating 484 authentic web pages from the Common Crawl dataset, constructing the first real-world benchmark for design-to-code evaluation. Building upon this foundation, WebCode2M~\cite{gui2025webcode2m} significantly expanded the scale with 20,000 samples, providing both comprehensive training data for model development and robust test sets for evaluation.  Other Benchmarks like Interaction2Code~\cite{xiao2024interaction2code} and MRWeb~\cite{wan2024mrweb} mainly focus on the interactive and multi-page web application generation, which is out of our scope.

% thereby establishing a more substantial foundation for advancing MLLM capabilities in web development tasks.

% To improve MLLM's UI-to-Code capabilities, pix2code define a Domain-specific language and leveraging CNN and LSTM to translate UI into DSL code. However, due to the limitations of DSL, pix2code has poor scalability. WebSight synthesized high-quality HTML code for training. To evaluate the MLLM’s success at HTML parsing, Web2Code~\cite{yun2024web2code}
% proposes the Webpage Code Generation Benchmark (WCGB).

% However, these two benchmarks are synthetic. Design2Code \cite{si2024design2code} manually select 484 real web pages from the Common Crawl dataset and constructed a real-world benchmark to evaluate the performance of MLLM on the design-to-code task. Compared to Design2Code, WebCode2M~\cite{gui2025webcode2m} with 20,000 samples is considerably larger, providing both a training set for model learning and a test set for evaluation.

Table~\ref{tab:bench_compare} presents a comparison between existing benchmarks and our, in terms of the size, collection method, framework, target tasks, and evaluation dimension. 
% While the aforementioned benchmarks have gained significant popularity among researchers for evaluating the performance of MLLMs on the front-end code generation task, 
\textbf{\bench~is distinct in incorporating varied front-end frameworks, multiple tasks, and diverse evaluation dimensions.}

\section{\bench}

% Fig.~\ref{fig:bench} shows the construction pipeline of \bench,

\subsection{Task Definition}

% We divide the process of automated front-end development into the following three tasks: ``Design Generation'', ``Design Edit'' and ``Design Repair'' as shown in Fig.~\ref{fig:task}. These three tasks simulate the three steps involved in UI development by front-end developers.

% As shown in Fig.~\ref{fig:task}, the automated front-end development mainly contains the following three tasks: ``Design Generation'', ``Design Edit'' and ``Design Repair''.

% \textbf{Design Generation (V-to-PL)}. The objective of design generation is to generate expected code based on the UI Mockups. The input contains the UI design image, the output is the UI code.

% \textbf{Design Edit (V\&NL\&PL-to-PL)}. The goal of the Design Edit task is to generate front-end code that complies with user modification instructions. The input contains the original UI design image, original UI code and user instruction which described by natural language.

% \textbf{Design Repair (V\&NL\&PL-to-PL)}. The goal of the design repair is to repair the UI code. Here, the issues contains the visual design issue and the compilation issues.

As shown in Fig.~\ref{fig:task}, the automated front-end development mainly contains the following three tasks: ``Design Generation'', ``Design Edit'' and ``Design Repair''.

\textbf{Design Generation ($\mathcal{T}_G$)}. The objective of design generation is to generate expected code based on the UI Mockups. Formally, given a UI design image $I$, the task aims to generate corresponding UI code $C$ such that $\mathcal{T}_G: I \rightarrow C$. The input contains the UI design image $I$, and the output is the UI code $C$ that accurately reproduces the visual layout and styling.

\textbf{Design Edit ($\mathcal{T}_E$)}. The goal of the design edit is to generate front-end code that complies with user modification instructions. Given the original UI design image $I_{o}$, original UI code $C_{o}$, and user instruction $T$ described in natural language, the task produces modified code $C_{new}$ such that $\mathcal{T}_E: (I_{o}, C_{o}, T) \rightarrow C_{new}$. The input contains the original UI design image $I_{o}$, original UI code $C_o$, and user instruction $T$, while the output is the updated code $C_{new}$ incorporating the requested modifications.

% \textbf{Design Repair ($\mathcal{T}_R$)}. The goal of the design repair is to repair the UI code with identified issues. Given the original UI code $C{orig}$ and error description $E$ (which may include visual design issues or compilation errors), the task generates corrected code $C_{fixed}$ such that $\mathcal{T}R: (C{o}, E) \rightarrow C_{fixed}$. The input contains the problematic UI code and error specifications, and the output is the repaired code that resolves both visual design issues and compilation problems.

\textbf{Design Repair ($\mathcal{T}_R$)}. The goal of the design repair is to repair the UI code with display issues. Given the problematic UI code $C_{p}$, the problematic UI image $I_{p}$, the task generates repaired UI code $C_{r}$ such that $\mathcal{T}_R: (C_{p}, I_{p}) \rightarrow C_{r}$. The input contains the problematic UI code $C_p$ and image $I_p$, the output is the repaired code $C_r$ that resolves visual design issues.

% and compilation problems.

% and error description $E$ (include visual design issues), 

\subsection{Data Collection}

\begin{figure}[t]
\centering
\includegraphics[width = .95\textwidth]{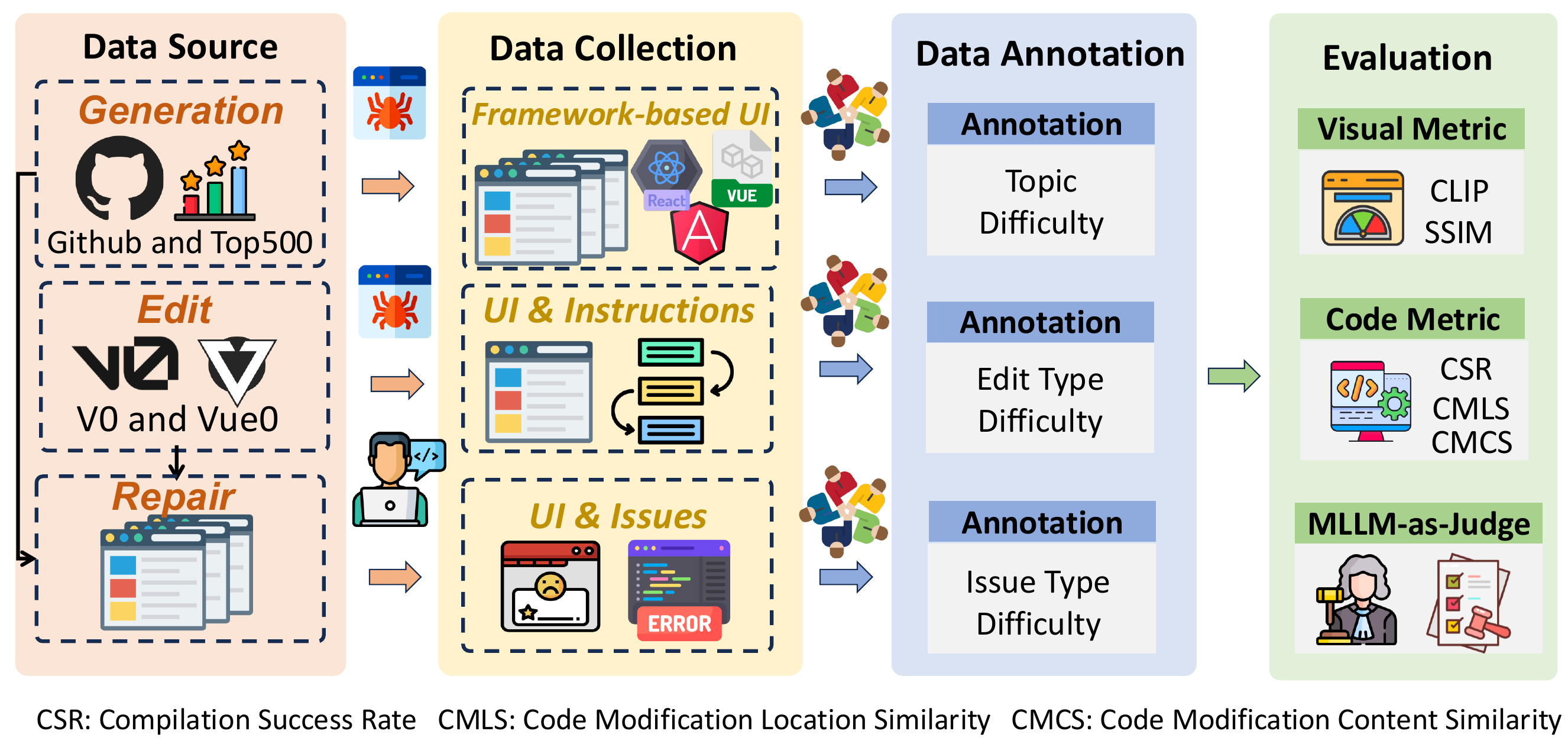}
\caption{Pipeline of \bench \  construction.}
\label{fig:bench}
\end{figure}

% \begin{figure}[ht]
% \centering
% \includegraphics[width = .48\textwidth]{figures/DesignBench1.pdf}
% \caption{Pipeline of \bench \  construction.}
% \label{fig:bench}
% \end{figure}

\textbf{Design Generation.} For websites developed by frameworks, we collect samples from GitHub and the top 500 globally visited websites. Existing reports \cite{web-framework, FrontendFrameworksPopularity}~show that the three most popular front-end development frameworks are React, Vue, and Angular, so we collect data for these three frameworks to build our benchmark. (1) Github projects. We search for ``React projects'', ``Vue projects'' and ``Angular projects'' to get a summary list of web projects with different frameworks, then we identify 152 popular projects with deployed links and higher star counts. These projects represent various real-world website uses, ranging from commercial product front-end websites to blogs, with 4055 average star counts and an average of 996 forks. Their popularity has proven their usefulness and quality. (2) Top 500 globally visited websites. These websites are ranked by Moz~\cite{top500}, and we collect 158 webpages that adopt frameworks as our dataset. 

For websites without applying frameworks, we sample 120 webpages with different lengths from the webcode2m dataset~\cite{gui2025webcode2m}. Ultimately, we compile a dataset consisting of 430 webpages. After getting the link of the webpage, we apply single-file tool~\cite{single-file-cli} to save the webpage as a stand-alone file, then replace the images in the web page with placeholders, and finally use selenium~\cite{selenium} to take screenshots for constructing the input design image $I$ for task $\mathcal{T}_G$.

\textbf{Design Edit.} To obtain real-world user instructions for UI design modification, we crawl 541 React-based projects from Vercel's V0 platform~\cite{v0dev} and 1,349 Vue-based projects from Vue0~\cite{vue0dev}. These projects contain comprehensive interaction histories, including user instructions and corresponding UI code implementations.

We first filter the projects that can compile successfully with at least two iteration rounds.  For each project with consecutive editing steps $v_{1}$ and $v_{2}$, we construct evaluation pairs where the input consists of the previous version's code $C_{v1}$, rendered UI image $I_{v1}$, and the user instruction $T_{v1 \rightarrow v2}$ that describes the desired modifications. The ground truth comprises the updated code $C_{v2}$ and its corresponding rendered UI image $I_{v2}$ that implements the requested changes.

However, some edit steps feature ambiguous user instructions or poorly UI modifications that fail to satisfy the intended requirements (here\footnote{https://github.com/WebPAI/DesignBench/blob/main/assets/Unclear\_Failed\_Examples.md} are some bad cases). To filter out low-quality samples, we employ five PhD students with three years of front-end development experience to conduct a comprehensive assessment.

Specifically, the annotators classify user instructions into three levels of clarity: \textit{clear}, \textit{moderate}, and \textit{ambiguous}, based on detailed assessment of instruction specificity, actionability, and comprehensibility. Similarly, they evaluate the quality of modified UI and categorize them into \textit{terrible}, \textit{good}, and \textit{excellent} based on how well the changes align with the given instructions. The detailed annotation guidelines are in~\cite{DesignBench}. The final classifications for both instruction clarity and UI modification effectiveness are determined through a majority voting process among the annotators.

% Finally, we compile 359 high quality React and Vue samples, to obtain Angular and native HTML/CSS samples, we randomly select 146 samples to translate them using MLLMs into angular and vanilla code. After the translation, five PhD students will check the correctness of the implementation and make necessary modifications.

Finally, we curate 359 high-quality React and Vue samples that receive \textit{clear} instruction ratings and \textit{excellent} modification quality scores. To obtain Angular and vanilla HTML/CSS samples, we randomly select 146 samples from this curated set and translate them into Angular and vanilla HTML/CSS code implementations using GPT-4o~\cite{openai_gpt4o}. Following the automated translation, the same five PhD students verify the correctness of the translated implementations, and make necessary modifications to ensure the code meet users' instructions.

\textbf{Design Repair.} After collecting the webpages of the above two tasks, the five PhD students filter the samples with design issues. They follow comprehensive guidelines to identify display issues, including: (1) \textit{Layout issues} such as misaligned elements and incorrect positioning; (2) \textit{Visual inconsistencies} including improper spacing, incorrect font sizes, or color scheme violations; (3) \textit{Component rendering problems} such as missing elements, overlapping content, or distorted images.

% he detailed annotation guidelines are in~\cite{DesignBench}.

Finally, we get 111 webpages with problematic UI code $C_p$ and image $I_p$. To obtain the repaired code $C_r$ for evaluation, 111 webpages are evenly assigned to the five PhD students to manually fix the issues. Each student spent about 5 hours to complete the repair of all assigned samples. The detailed annotation guidelines of these three tasks are in our repo\footnote{\url{https://github.com/WebPAI/DesignBench/blob/main/assets/Guideline.md}}.

% (4) \textit{Interactive element failures} including non-functional buttons, broken navigation, or inaccessible form controls; and (5) \textit{Cross-browser compatibility issues} that affect rendering consistency. Each UI sample is evaluated against these criteria, and issues are documented with specific descriptions to guide the repair process. Only samples with clear, actionable design issues are included in the repair dataset to ensure meaningful evaluation of model debugging capabilities.

% We first screened out problematic UIs from the datasets of the Design Generation and Design Edit tasks.

% We employ four PhD students with three years of front-end development experience are employed to annotate the UI issues

\subsection{Data Annotation}
\label{subsec:da}

\begin{figure*}[ht]
\centering
\vspace{-0.1in}
\includegraphics[width = .99\textwidth]{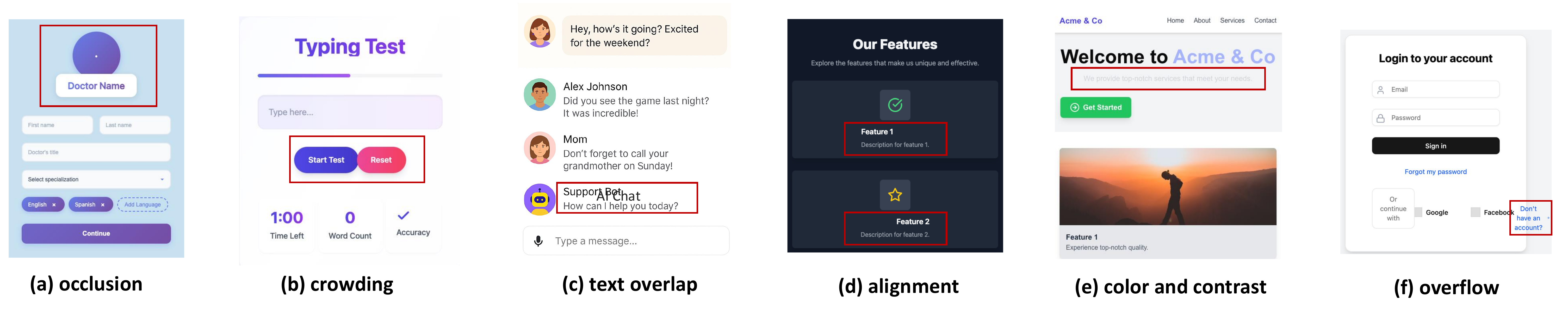}
\vspace{-0.1in}
\caption{The UI issue types in \bench. The red bounding box marks the issue location.}
\vspace{-0.2in}
\label{fig:repair}
\end{figure*}

% \subsubsection{Design Generation Task}. For websites developed by frameworks, we collect from github and top 500 globally visited websites. (1) Github projects. We search for ``React projects'', ``Vue projects'' and ``Angular projects'' to get a summary list of web projects with different frameworks, then we identify xxxxx popular projects with deployed links and higher star counts. These projects represent various real-world website uses, ranging from commercial product front-end websites to blogs, with xxxx average star counts. Their popularity have proven their usefulness and quality. (2) Top 500 globally visited websites, these websites are as ranked by Moz \footnote{https://moz.com/top500}, we collect the websites that adopt frameworks as our dataset. 

% For websites without applying frameworks, we sample 120 webpages with different lengths from the webcode2m dataset~\cite{gui2025webcode2m} . Ultimately, we compile a dataset consisting of 430 webpages.

% \subsubsection{Design Edit Task}. To obtain real user prompts for modifying the UI design, we crawl 541 and 1349 projects from Vercel’s V0 projects developed by React \footnote{https://v0.dev/} and Vue0 projects \footnote{https://www.vue0.dev/} developed by Vue respectively.

\textbf{Design Generation.} The annotators are employed to annotate the topics of the webpages based on their functions.

\textbf{Design Edit.} The annotators are instructed to annotate both the type and difficulty level of UI modifications. The modification type encompasses two dimensions: the operation type and the corresponding UI attribute type being adjusted. The operation types include three fundamental categories: Add (introducing new UI elements), Change (modifying elements), and Delete (removing elements). The UI attribute types subject to adjustment comprise six main categories: text (including content, font, and typography modifications), color (encompassing background colors, text colors, and accent colors), position (spatial arrangement and layout adjustments), size (dimensional scaling and resizing operations), shape (geometric modifications and structural changes), and component-level (holistic modifications affecting entire UI components). The difficulty level reflects several factors, primarily the number of UI elements that require adjustment, the complexity of the interdependencies between modified elements, and the scope of cascading changes required to maintain design consistency and functionality throughout the interface.

\textbf{Design Repair.} The annotators are employed to annotate the UI issues. After collecting the webpages with display issues,  we first randomly select 25\% samples for analysis and then discuss, revise, and refine the UI issue type until everyone reaches a consensus. During annotating new data, if encountering a new issue type, annotators will communicate and update issue type in time to guide subsequent annotations. Finally, we classify 6 types of UI display issues as follows:

\begin{itemize}[leftmargin=*]
    \item \textit{Occlusion.} Elements are hidden or partially covered by other elements, making content inaccessible or invisible to users. This includes overlapping components, modal dialogs blocking content, or elements positioned behind others. As shown in Fig.~\ref{fig:repair}(a). The ``Doctor Name'' box partially covers the portrait frame.
    
    % making content inaccessible or invisible to users. This includes overlapping components, modal dialogs blocking content, or elements positioned behind others.
    
    \item \textit{Crowding.} Too many elements are packed into a small space without adequate spacing, making the interface feel cluttered and difficult to navigate. Fig.~\ref{fig:repair}(b) shows an example that the ``Start Test'' and ``Reset'' buttons are tightly packed together.

    \item \textit{Text overlap.} Text content overlaps with other text or UI elements, making it unreadable or causing visual confusion. Fig.~\ref{fig:repair}(c) shows that the ``AI Chat'' text and ``Support Bot'' text are overlapped.

    \item \textit{Alignment.} Elements are not properly aligned with each other or the overall layout grid, creating a disorganized appearance. As shown in Fig.~\ref{fig:repair}(d), the ``Feature 1'' and ``Feature 2'' titles are not aligned with icons.
    
    \item \textit{Color and contrast.} Poor color choices that affect readability or accessibility, including insufficient contrast between text and background, or color combinations that are difficult for users with visual impairments to distinguish. As illustrated in Fig.~\ref{fig:repair}(e), the text color and background color are too close to each other, making it difficult to distinguish.
    \item \textit{Overflow.} Content extends beyond its intended container boundaries, causing horizontal scrollbars, cut-off text, or elements appearing outside their designated areas. For example, the text ``Don't have an account'' exceeds the login container.
\end{itemize}

% \subsection{Difficulty Classification}

% We divide different samples into different levels of difficulty to measure the performance of MLLM on samples of different difficulty levels.

% However, the difficulty measurement indicators of the three tasks are different.

% The difficulty levels for each task are determined by distinct factors that reflect the inherent challenges of front-end development scenarios. 
% and scope of the required modifications.

% For Design Generation task, which require generating code from scratch, difficulty is primarily measured by the visual complexity of the target webpage.

% In Design Edit task, where models must modify existing code according to user instructions, difficulty correlates with the complexity of users' instructions.

% For Design Repair tasks, difficulty is assessed based on the severity of UI issues, specifically quantified by the extent of code modifications necessary to resolve the identified problems.

% from minor styling fixes to substantial architectural corrections affecting multiple code segments.

\subsection{Benchmark Statistics}

Table~\ref{tab:dis} presents the sample counts of the three tasks with three frameworks, the Vanilla denotes the webpage developed by vanilla HTML/CSS.

\begin{table}[h]
\centering
\caption{Sample counts of three frameworks on three tasks.}
\label{tab:dis}
\begin{tabular}{@{}c|ccc@{}}
\toprule
Framework & Design Generation & DesignEdit & Design Repair \\ \midrule
React     & 109               & 108        & 28            \\
Vue       & 118               & 105        & 27            \\
Angular   & 83                & 66         & 28            \\
Vanilla   & 120               & 80         & 28            \\ \hline
Total     & 430               & 359        & 111           \\ \bottomrule
\end{tabular}
\end{table}

% \begin{figure}[t]
%     \subfigure[Topic Distribution.]{
%     \label{fig:topic}
%     \centering
%     \includegraphics[width = .42\textwidth]{figures/topic.pdf}
%     }
%     \subfigure[Issue Distribution.]{
%     \label{fig:issue}
%     \centering
%     \includegraphics[width = .42\textwidth]{figures/issue.pdf}
%     }
%     \caption{Category distribution of Design Generation and Design Repair task.}
%     \vspace{-0.1in}
% \end{figure}

% \begin{figure}[t]
%     \subfigure[Operation Distribution.]{
%     \label{fig:operation}
%     \centering
%     \includegraphics[width = .41\textwidth]{figures/operation.pdf}
%     }
%     \subfigure[Visual Distribution.]{
%     \label{fig:visual}
%     \centering
%     \includegraphics[width = .42\textwidth]{figures/visual.pdf}
%     }
%     \caption{Edit type distribution of Design Edit task.}
%     \vspace{-0.1in}
%     \label{fig:edit}
% \end{figure}

\begin{figure}[t]
    \subfigure[Topic Distribution.]{
    \label{fig:topic}
    \centering
    \includegraphics[width = .22\textwidth]{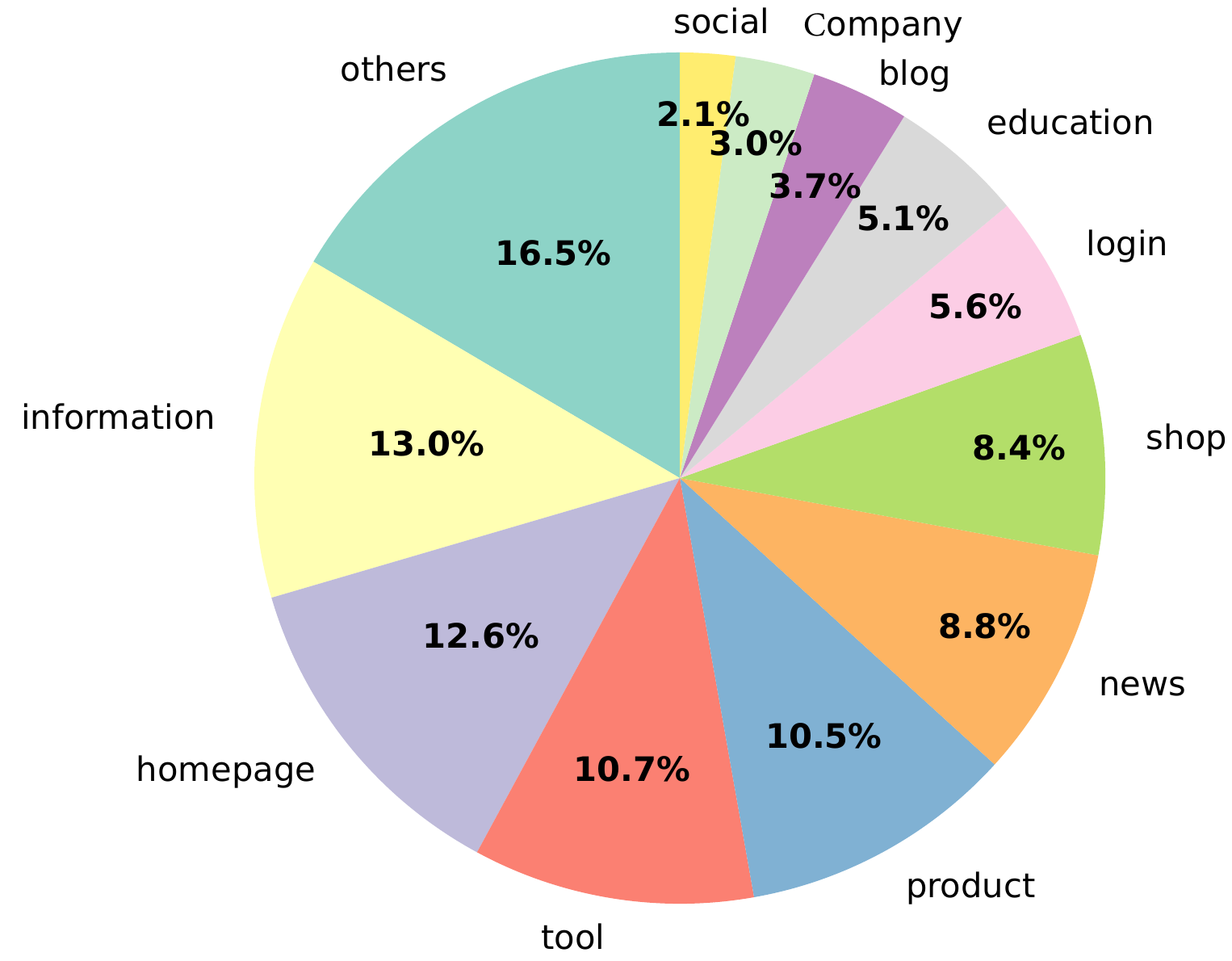}
    }
    % \caption{Category distribution of Design Generation and Design Repair task.}
    \subfigure[Operation Distribution.]{
    \label{fig:operation}
    \centering
    \includegraphics[width = .21\textwidth]{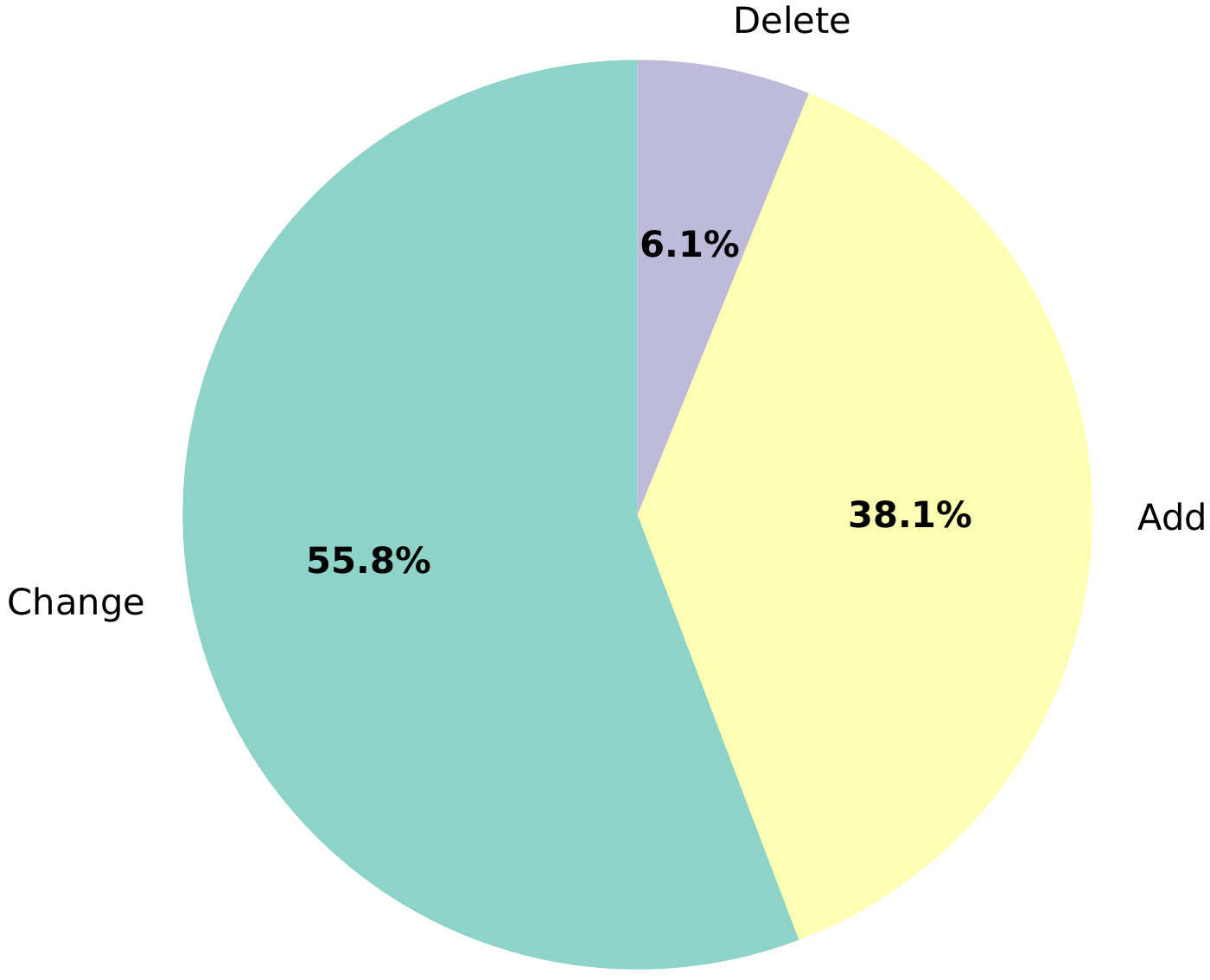}
    }
    \subfigure[Visual Distribution.]{
    \label{fig:visual}
    \centering
    \includegraphics[width = .22\textwidth]{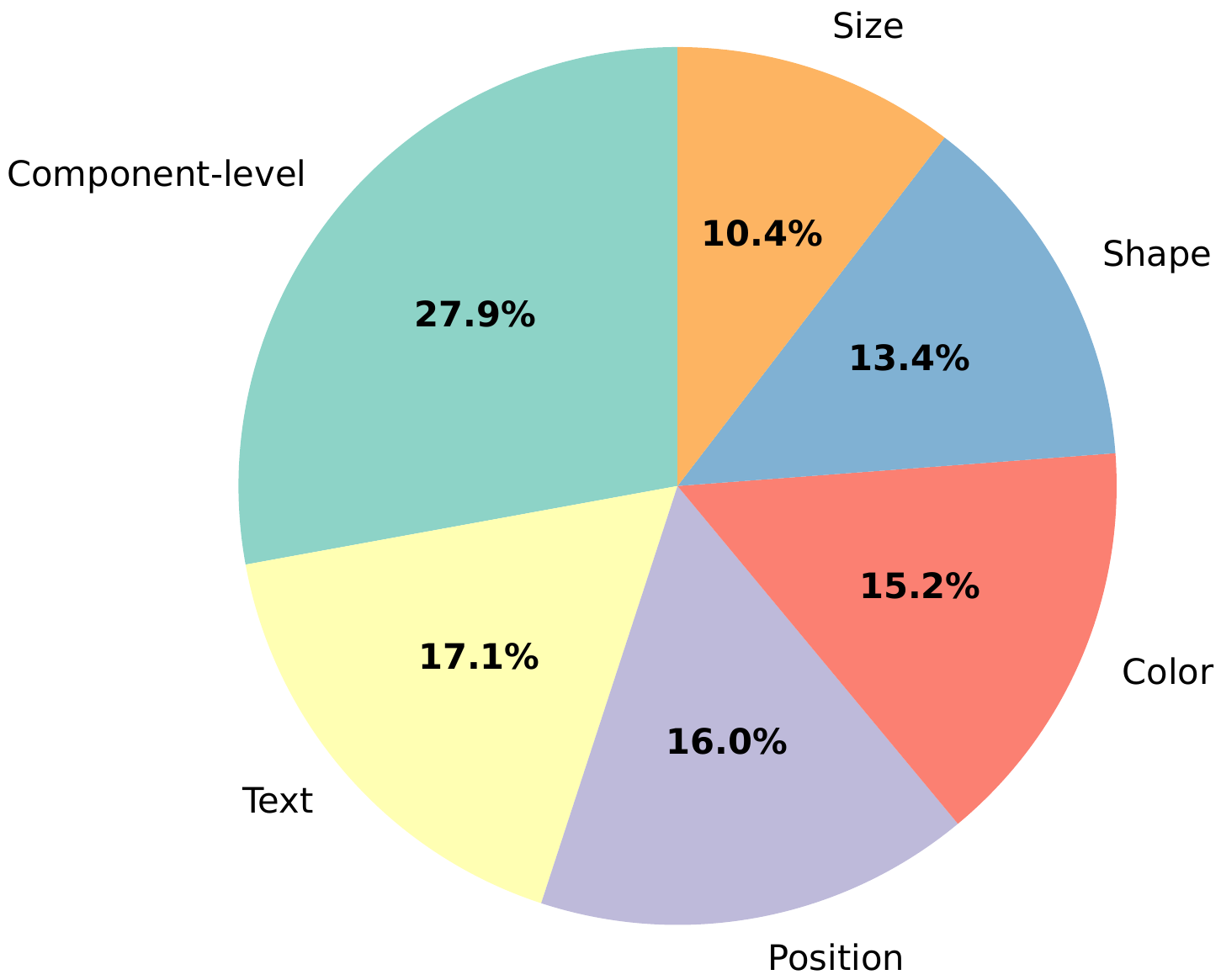}
    }
    \subfigure[Issue Distribution.]{
    \label{fig:issue}
    \centering
    \includegraphics[width = .22\textwidth]{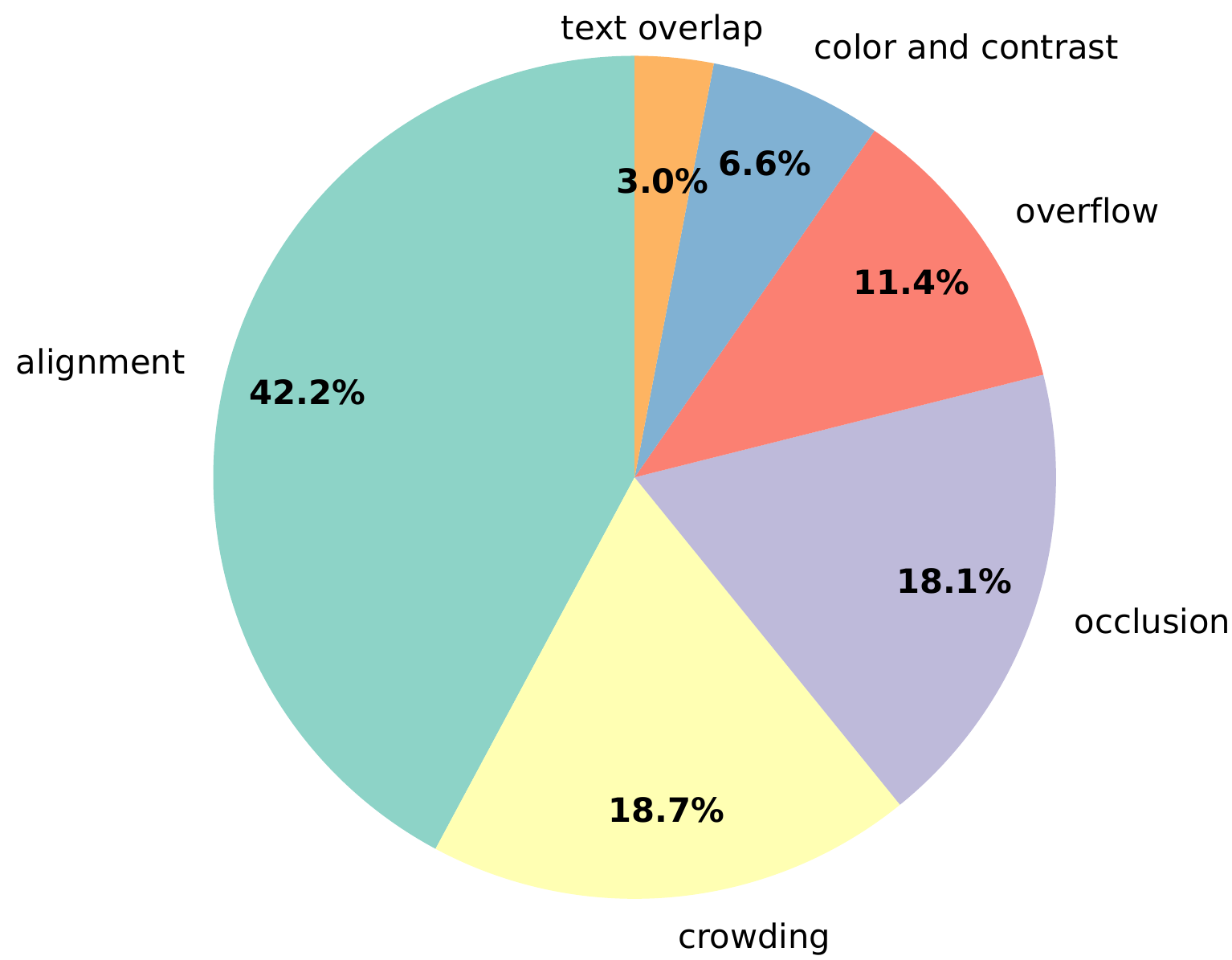}
    }
    \caption{Type distribution of three tasks.}
    \label{fig:edit}
\end{figure}

% \begin{figure}[t]
%     \subfigure[Operation Distribution.]{
%     \label{fig:operation}
%     \centering
%     \includegraphics[width = .41\textwidth]{figures/operation.pdf}
%     }
%     \subfigure[Visual Distribution.]{
%     \label{fig:visual}
%     \centering
%     \includegraphics[width = .42\textwidth]{figures/visual.pdf}
%     }
%     \caption{Edit type distribution of Design Edit task.}
%     \vspace{-0.1in}
%     \label{fig:edit}
% \end{figure}

Fig.~\ref{fig:topic} shows that \bench \ covers a diverse range of web topics with more than 11 types, including information, homepage, tool, product, news, and so on. This extensive topic coverage demonstrates that \bench \ encompasses a broad range of aspects of digital interface design, ensuring comprehensive evaluation across different domain requirements.

Fig.~\ref{fig:edit} shows the distribution of the edit type of design edit tasks. The operation distribution reveals three main categories: Change (55.8\%), Add (38.1\%), and Delete (6.1\%). The corresponding visual modifications encompass diverse types, including component-level changes (27.9\%), text modifications (17.1\%), position adjustments (16.0\%), color alterations (15.2\%), shape modifications (13.4\%), and size adjustments (10.4\%), indicating that design edit tasks cover a broad spectrum of modification requirements.

Fig.~\ref{fig:issue} shows the distribution of the UI issues. The issue types span multiple dimensions with alignment being most prevalent (42.2\%), followed by crowding (18.7\%), occlusion (18.1\%), overflow (11.4\%), and other visual defects. This wide range of issue types demonstrates that design repair tasks address comprehensive quality assurance across multiple design issues dimensions.

\section{Experiment Setup}

\subsection{Prompt Design}

\begin{figure*}[ht]
\centering
\includegraphics[width = .99\textwidth]{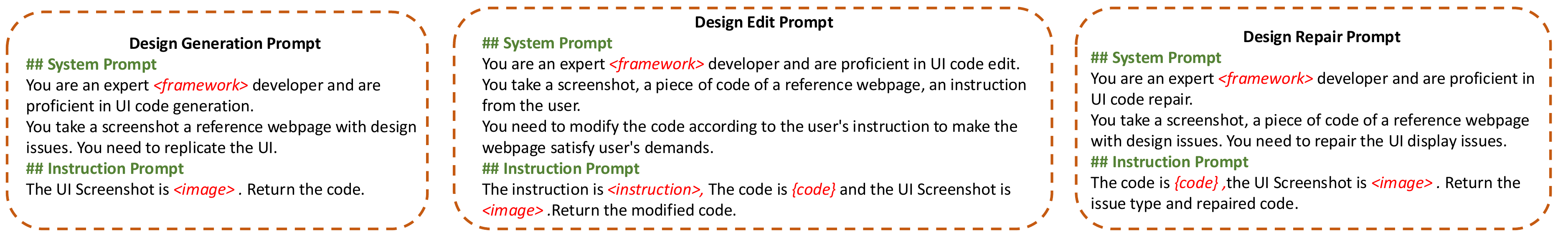}
\caption{Prompts of design generation, edit and repair task.}
\label{fig:prompt}
\end{figure*}

The simplified prompts for the three tasks are illustrated in Fig.\ref{fig:prompt}, with detailed prompts available in our code~\cite{DesignBench}. Our prompts are designed with the following structure: (1) background description: a concise scene setting statement (“You are a senior vue/react/angular developer", "You are tasked with generating/editting/repairing") primes the model for the correct role and task description; (2) detailed and specific task requirements: For design generation, exactly reproduce the screenshot: fonts, colors, layout, spacing. For design edit, surgically alter only the elements specified in the  natural-language instruction while leaving all other visuals untouched. For design repair, identify and correct the listed UI defects (occlusion, crowding, overlap, alignment, color/contrast, overflow). (3) framework-specific coding conventions: the prompts guide the model to import available UI components and follow framework-specific coding conventions, with a generic example to reduce ambiguity; (4) output format instructions and examples, each prompt includes strict output format requirements to ensure that the generated code can be reliably extracted and evaluated.

\subsection{Models}

The studied MLLMs are listed in Table~\ref{table:models}. We select six state-of-the-art LLMs that have been widely explored in multimodal tasks, three from open-source models, namely Pixtral~\cite{agrawal2024pixtral}, Qwen~\cite{Qwen2.5-VL}, and LLama~\cite{meta_llama}, three from commercial models like Gemini~\cite{google_gemini_api}, GPT~\cite{openai_gpt4o} and Claude~\cite{anthropic_claude}. In configuring the MLLM models, we set the temperature to 0 and the maximum number of tokens output as the upper limit of MLLMs' maximum output token. All other parameters are kept at their default settings. The detailed prompts of the three tasks are available in our code repository~\cite{DesignBench}. The entire benchmark evaluation incurs an average API cost of \$52 per model on average. The average processing times per sample with a single thread, are 49 seconds for generation, 29 seconds for editing, and 25 seconds for repair, respectively.

\subsection{Data Contamination Check}
\label{subsec:contamination}

First, in terms of the data source, we review the technical reports and training datasets of open source MLLMs. The training datasets for Qwen~\cite{Qwen2.5-VL}, Pixtral~\cite{agrawal2024pixtral} and LLama~\cite{meta_llama} includes image captions, interleaved image-text data, optical character recognition (OCR) data, visual knowledge, crawling plain text and web document for Q\&A, but not web html code. Furthermore, data leakage is particularly unlikely for design edit and design repair, because for design edit task, the training datasets don't contain ``<instruction, image, react/vue/angular/html code>'' triple corpora. For design repair, there is no ground-truth repair labels are publicly available, as the benchmark is generated and verified by our annotators.
To confirm that MLLMs are not ``cheating'' by memorizing and reproducing the original webpage code, we measure the BLEU score~\cite{papineni2002bleu} between the generated and original webpage code in the following table. The low BLEU score indicates that the model generates code from the design images rather than directly copying the original webpage.

% Although there is decent visual similarity, the low BLEU score indicates that the model generates code from the design images rather than directly copying the original webpage.

% \begin{table}[htbp]
% \centering
% % \renewcommand{\arraystretch}{1.5}
% % \setlength{\tabcolsep}{10pt}
% % \arrayrulecolor[HTML]{D3D3F5}
% \begin{tabular}{|c|c|c|c|c|c|c|c|c|}
% \hline
% \textbf{Claude-3.7} & \textbf{GPT-4o} & \textbf{Gemini-2.0} & \textbf{Llama-90B} & \textbf{Llama-11B} & \textbf{Pixtral-124B} & \textbf{Pixtral-12B} & \textbf{Qwen-72B} & \textbf{Qwen-7B} \\ \hline
% 0.1531 & 0.1318 & 0.1200 & 0.0771 & 0.0562 & 0.1117 & 0.1151 & 0.1338 & 0.0907 \\ \hline
% \end{tabular}
% \caption{Performance Comparison}
% \label{tab:performance_comparison}
% \end{table}

\begin{table}[htbp]
\centering
\setlength{\tabcolsep}{0.1em}
\caption{The BLEU score between original webpage code and generated code.}
\label{table:contamination}
\resizebox{\textwidth}{!}{
\begin{tabular}{ccccccccc}
\toprule
Claude-3.7 & GPT-4o & Gemini-2.0 & Llama-90B & Llama-11B & Pixtral-124B & Pixtral-12B & Qwen-72B & Qwen-7B \\
\midrule
0.1531 & 0.1318 & 0.1200 & 0.0771 & 0.0562 & 0.1117 & 0.1151 & 0.1338 & 0.0907 \\
\bottomrule
\end{tabular}}
\end{table}

% Although the C4 dataset is widely used for training large models, it is unlikely that our test set was seen by MLLMs during training. The C4 dataset contains only text data, without HTML/CSS/JavaScript code, and current MLLM training datasets, like Qwen~\cite{Qwen2.5-VL}, do not include interactive prototyping data. Additionally, many existing benchmarks~\cite{si2024design2code, gui2024vision2ui} derived from the C4 dataset still present challenges for MLLMs. To confirm that MLLMs are not ``cheating'' by recognizing and reciting the original webpage code, we compute the BLEU~\cite{papineni2002bleu} score between the generated and original webpage HTML. The very low BLEU score (Table~\ref{tab:contamination}) indicates that the model generates the code based on the interactive prototype, not by directly copying the original webpage.

% The API cost for all the models are approximate 470 dollars. The time cost is about

% The prompts for the three tasks are illustrated in Fig.\ref{fig:prompt}, with detailed prompts available in our code repository\cite{DesignBench}.

% for Pixtral, LLaMa, Gemini, GPT and Claude as 8192. For the Qwen series models, the maximum output token are set to 2048. All other parameters were kept at their default settings.

% as outlined in the relevant API documentation.
% code generation tasks.

\subsection{Metric}
\label{subsec:metric}

% We evaluate the generation code using
% three progressive metrics: completion/compilation/test case pass
% rate.

% We evaluate the performance of the model on \bench \ from three types of metric: (1) visual metric: (2) code metric: (3) score metric:

% We evaluate the performance of the model on \bench \ from three types of metrics:

\textbf{Visual Metrics}. (1) \textbf{CLIP}~\cite{radford2021learning} is applied to  measure the semantic similarity between the generated and original webpages. (2) \textbf{SSIM}~\cite{wang2004image} (Structural Similarity Index Measure) score is applied to calculate the structure similarity. It evaluates the layout and compositional accuracy, emphasizing the spatial arrangement and structural similarities between the generated and original webpages.

% \textbf{Code Metrics}. (1) \textbf{Compilation Success Rate (CSR)}. This metric represents the percentage of generated code that compiles successfully without errors. Assume that the total number of samples is N and the number of samples compiled successfully is S, then $CSR=\frac{S}{N}$. (2) \textbf{Code Modification Similarity (CMS)}. We employ the Jaccard similarity~\cite{thada2013jaccard} to quantify the precision of code modifications on design edit and design repair tasks by comparing the sets of modified line numbers between the ground truth and generated code. Let $A$ represent the set of line numbers modified in the ground truth code and $B$ represent the set of line numbers modified in the generated code. The CMS is formally defined as: $CMS(A, B) = \frac{|A \cap B|}{|A \cup B|}$.

\textbf{Code Metrics}. (1) \textbf{Compilation Success Rate (CSR)}. This metric represents the percentage of generated code that compiles successfully without errors. Assume that the total number of samples is N and the number of samples compiled successfully is S, then $CSR=\frac{S}{N}$. (2) \textbf{Code Modification Location Similarity (CMLS)}. We employ the Jaccard similarity~\cite{thada2013jaccard} to quantify the precision of code modifications on design edit and design repair tasks by comparing the sets of code modification operations between the ground truth and generated code. We first parse the code into an Abstract Syntax Tree (AST), and then identify the operation (i.e., add, delete and modify) by the AST node comparison. Let $A$ represent the set of code modification operations in the ground truth code and $B$ represent the set of code modification operations in the generated code. The CMLS is formally defined as: $CMLS(A, B) = \frac{|A \cap B|}{|A \cup B|}$. (3) \textbf{Code Modification Content Similarity (CMCS).} 
While CMLS focuses on modification locations, CMCS evaluates the semantic correctness of modified code content. Given modification operation sets $A$ and $B$, we match operations based on AST node correspondence and denote the matched pairs as $M \subseteq A \times B$. For each $(a,b)\in M$, CodeBLEU~\cite{ren2020codebleu} is used to compute content similarity, while unmatched operations receive a score of zero. The CMCS is defined as: $CMCS(A,B) = \frac{1}{|A \cup B|} \sum_{(a,b)\in M} \text{CodeBLEU}(a,b)$.

\textbf{MLLM-as-Judge Metrics.} MLLMs have shown great performance in assisting judges across diverse modalities~\cite{chen2024mllm, wang2025can}. Therefore, we prompt GPT-4o~\cite{openai_gpt4o} to determine whether the model meets the user's requirements on the design edit task and resolve the design issues on the design repair task, and output an \textbf{MLLM score} between 0 and 10 with detailed explanations (0-3 denotes the poor edit/repair, 4-6 denotes partial edit/repair, 7-8 denotes Good edit/repair and 9-10 denotes excellent edit/repair). The detailed judge prompts can be found in our repo\footnote{https://github.com/WebPAI/DesignBench/blob/main/code/evaluator/metric.py}. We employ the same five PhD students to conduct human validation studies to verify the reliability and consistency of our MLLM-as-Judge approach. Final validation are determined by majority voting. The Kappa score for inter-annotator agreement is 0.8648 and 0.8428 on design edit and repair task respectively, indicating "almost perfect" consensus. For design edit task, we sample 359 samples and validate this MLLM score through human evaluation, which can achieve an average accuracy of \textbf{95.54\%}. For design repair task,  we sample 111 samples and validate this MLLM score through human evaluation, which can achieve an average accuracy of \textbf{91.89\%}.

\begin{table}
\caption{Studied Multimodal Large Language Models.}
\label{table:models}
\centering
\resizebox{0.6\textwidth}{!}{
\begin{tabular}{l|l|l|l}
\toprule
Base Model & Model & Abbreviation & Size \\
\midrule
Pixtral~\cite{agrawal2024pixtral} & Pixtral-12B-2409 & Pixtral-12B & 12B\\
Pixtral~\cite{agrawal2024pixtral} & Pixtral-large-latest & Pixtral-124B & 124B \\
Qwen~\cite{Qwen2.5-VL} & Qwen2.5-VL-7B-Instruct & Qwen-7B & 7B\\
Qwen~\cite{Qwen2.5-VL} & Qwen2.5-VL-72B-Instruct & Qwen-72B & 72B \\
LLama~\cite{meta-llama} & Llama-3.2-11B-Vision-Instruct & Llama-11B & 11B  \\
LLama~\cite{meta-llama} & Llama-3.2-90B-Vision-Instruct & LLama-90B & 90B \\
Gemini~\cite{google_gemini_api} & Gemini-2.0-Flash & Gemini-2.0 & -  \\
GPT~\cite{openai_gpt4o} & GPT-4o-2024-11-20 & GPT-4o & -  \\
Claude~\cite{anthropic_claude} & Claude-3-7-sonnet-20250219 & Claude-3.7 & -\\
\bottomrule
\end{tabular}}
\vspace{-0.1in}
\end{table}

\subsection{Research Questions}

% The research questions (RQs) are designed as follows:

\begin{itemize}[leftmargin=*]
    \item \textbf{RQ1:} (Performance across tasks) How do MLLMs perform across distinct front-end tasks?
    \item \textbf{RQ2:} (Performance across frameworks) What is the comparative performance of MLLMs when applied to different development frameworks?
    \item \textbf{RQ3:} (Difficulty Influence) How does varying task difficulty impact MLLM performance?
    \item \textbf{RQ4:} (Context Influence) To what extent do different input contexts affect MLLMs' performance?
    \item \textbf{RQ5:} (Limitation analysis) What are the primary limitations of MLLMs in developing framework-based webpages?
    \item \textbf{RQ6:} (Failure analysis) What mistakes do MLLMs make on the three tasks?
\end{itemize}

\section{Experiment Results}

\begin{table*}[ht]
\centering
\setlength{\tabcolsep}{0.1em}
\caption{The model performance on \bench \ under different tasks and different frameworks. Bold numbers on a dark red background indicate the maximum values, and an underline with a light red background denotes the second-best value.}
\label{tab:overall}
\resizebox{0.95\textwidth}{!}{
\begin{tabular}{@{}c|c|ccc|cc|cc|cc@{}}
\toprule
\multirow{2}{*}{\bf Metric} & \multirow{2}{*}{\bf Framework} & \multicolumn{1}{c}{Claude} & \multicolumn{1}{c}{GPT} & \multicolumn{1}{c}{Gemini} & \multicolumn{2}{c}{Llama} & \multicolumn{2}{c}{Pixtral} & \multicolumn{2}{c}{Qwen} \\
\cmidrule(lr){3-3} \cmidrule(lr){4-4} \cmidrule(lr){5-5} \cmidrule(lr){6-7} \cmidrule(lr){8-9} \cmidrule(lr){10-11}
& & Claude-3.7 & GPT-4o & Gemini-2.0 & Llama-90B & Llama-11B & Pixtral-124B & Pixtral-12B & Qwen-72B & Qwen-7B \\
\hline
\multicolumn{11}{c}{\cellcolor{color2}\it Design Generation} \\
\hline
\multirow{4}{*}{\bf CLIP} & React     & \cellcolor{myred!100}{\textbf{0.8083}} & 0.7637 & 0.7611 & 0.7040 & 0.6401 & 0.7395 & 0.6168 & \cellcolor{myred!50}{\underline{0.7790}} & 0.0875 \\
& Vue       & \cellcolor{myred!100}{\textbf{0.8319}} & \cellcolor{myred!50}{\underline{0.7734}} & 0.6897 & 0.5323 & 0.3243 & 0.7811 & 0.7434 & 0.6836 & 0.0452 \\
& Angular   & \cellcolor{myred!50}{\underline{0.6024}} & 0.5963 & 0.6006 & 0.5327 & 0.4891 & \cellcolor{myred!100}{\textbf{0.6324}} & 0.4876 & 0.5149 & 0.0851 \\
& Vanilla   & \cellcolor{myred!100}{\textbf{0.8132}} & \cellcolor{myred!50}{\underline{0.7683}} & 0.7588 & 0.6404 & 0.6304 & 0.7403 & 0.7043 & 0.7597 & 0.7411 \\
\hline
\multirow{4}{*}{\bf SSIM} & React     & \cellcolor{myred!100}{\textbf{0.6912}} & 0.6655 & 0.6577 & 0.6463 & 0.6255 & 0.6478 & 0.5576 & \cellcolor{myred!50}{\underline{0.6774}} & 0.0832 \\
& Vue       & \cellcolor{myred!100}{\textbf{0.7155}} & \cellcolor{myred!50}{\underline{0.6809}} & 0.6136 & 0.4228 & 0.2727 & 0.6339 & 0.6432 & 0.5681 & 0.0411 \\
& Angular   & 0.4974 & 0.5196 & \cellcolor{myred!100}{\textbf{0.5251}} & 0.4883 & 0.4930 & \cellcolor{myred!50}{\underline{0.5138}} & 0.3996 & 0.4321 & 0.0814 \\
& Vanilla   & \cellcolor{myred!100}{\textbf{0.7004}} & 0.6788 & \cellcolor{myred!50}{\underline{0.6820}} & 0.6346 & 0.6687 & 0.6471 & 0.6229 & 0.6673 & 0.6476 \\
\hline
\multirow{4}{*}{\bf CSR} & React     & 0.9541 & \cellcolor{myred!100}{\textbf{0.9725}} & 0.9083 & 0.9450 & 0.8991 & \cellcolor{myred!100}{\textbf{0.9725}} & 0.8532 & 0.9541 & 0.1284 \\
& Vue       & \cellcolor{myred!100}{\textbf{0.9746}} & 0.9492 & 0.8390 & 0.7458 & 0.4915 & \cellcolor{myred!100}{\textbf{0.9746}} & 0.9407 & 0.8559 & 0.0678 \\
& Angular   & 0.6867 & 0.7108 & 0.7108 & \cellcolor{myred!50}{\underline{0.7349}} & 0.6988 & \cellcolor{myred!100}{\textbf{0.7590}} & 0.6024 & 0.6265 & 0.1205 \\
\hline
\multicolumn{11}{c}{\cellcolor{color2}\it Design Edit} \\
\hline
\multirow{4}{*}{\bf MLLM Score} & React     & \cellcolor{myred!100}{\textbf{8.1759}} & 8.0093 & 7.8148 & 6.1574 & 4.8148 & 8.0185 & 7.6111 & \cellcolor{myred!50}{\underline{8.0833}} & 1.8796 \\
& Vue       & \cellcolor{myred!100}{\textbf{8.3619}} & \cellcolor{myred!50}{\underline{8.1810}} & 8.0571 & 6.2571 & 3.1333 & 8.0381 & 7.0190 & 7.5714 & 2.2952 \\
& Angular   & 8.0152 & 8.2879 & \cellcolor{myred!100}{\textbf{9.1364}} & 5.6515 & 5.1212 & \cellcolor{myred!50}{\underline{8.6818}} & 7.8030 & 8.1970 & 2.0152 \\
& Vanilla  &  \cellcolor{myred!50}{\underline{9.1500}} & \cellcolor{myred!100}{\textbf{9.2250}} & 9.0250 & 7.7000 & 6.5750 & 9.1125 & 8.6000 & 9.1250 & 5.9125 \\

% \hline
% \multirow{4}{*}{\bf CMS}    & Vue       & \cellcolor{myred!100}{\textbf{0.4050}} & \cellcolor{myred!50}{\underline{0.3698}} & 0.3276 & 0.2104 & 0.0655 & 0.3046 & 0.2394 & 0.3276 & 0.0862 \\
% & React     & \cellcolor{myred!50}{\underline{0.4659}} & \cellcolor{myred!100}{\textbf{0.5246}} & 0.3710 & 0.2637 & 0.1819 & 0.4093 & 0.4120 & 0.4398 & 0.0815 \\
% & Angular   & \cellcolor{myred!100}{\textbf{0.6829}} & 0.6099 & \cellcolor{myred!50}{\underline{0.6392}} & 0.4700 & 0.3621 & 0.3264 & 0.2867 & 0.6018 & 0.1367 \\
% & Vanilla   &   \cellcolor{myred!50}{\underline{0.3439}} & 0.3394 & 0.2905 & 0.1946 & 0.1582 & \cellcolor{myred!100}{\textbf{0.3651}} & 0.2770 & 0.3209 & 0.1635 \\

\hline
\multirow{3}{*}{\bf CSR} & React & \cellcolor{myred!100}{\textbf{1.0000}} & 0.9815 & \cellcolor{myred!100}{\textbf{1.0000}} & 0.9167 & 0.7963 & 0.9907 & \cellcolor{myred!100}{\textbf{1.0000}} & 0.9907 & 0.4815 \\
& Vue  & \cellcolor{myred!100}{\textbf{0.9810}} & 0.9429 & 0.9524 & 0.9143 & 0.5905 & \cellcolor{myred!50}{\underline{0.9619}} & 0.9048 & 0.9333 & 0.4286 \\
& Angular  & 0.9091 & 0.9091 & \cellcolor{myred!100}{\textbf{1.0000}} & 0.8636 & 0.7727 & \cellcolor{myred!50}{\underline{0.9848}} & 0.9242 & 0.9091 & 0.3333 \\

\hline
\multirow{4}{*}{\bf CMLS}    & React       & 0.5555 & \cellcolor{myred!100}{\textbf{0.6159}} & 0.5325 & 0.3836  & 0.3272 & 0.5306 & 0.5209 & \cellcolor{myred!50}{\underline{0.5930}} & 0.1890 \\
& Vue     & \cellcolor{myred!100}{\textbf{0.5961}} & 0.5090 & \cellcolor{myred!50}{\underline{0.5414}} & 0.1924 & 0.1443 & 0.5290 & 0.3990 & 0.4720 & 0.1705 \\
& Angular   & \cellcolor{myred!100}{\textbf{0.7696}} & \cellcolor{myred!50}{\underline{0.7267}} & 0.6838 & 0.5057 & 0.4041 & 0.5176 & 0.5933 & 0.7016 & 0.3453 \\
& Vanilla   &   \cellcolor{myred!50}{\underline{0.5862}} & 0.5727 & 0.5645 & 0.3865 & 0.3113 & 0.5644 & 0.4764 & \cellcolor{myred!100}{\textbf{0.5782}} & 0.3125 \\

\hline
\multirow{4}{*}{\bf CMCS}    & React       & 0.4551 & \cellcolor{myred!100}{\textbf{0.5122}} & 0.4058 & 0.3022  & 0.2425 & 0.4149 & 0.4189 & \cellcolor{myred!50}{\underline{0.4872}} & 0.1319 \\
& Vue     & \cellcolor{myred!100}{\textbf{0.4807}} & 4232 & \cellcolor{myred!50}{\underline{0.4292}} & 0.1521 & 0.1016 & 0.4168 & 0.3199 & 0.3837 & 0.1139 \\
& Angular   & \cellcolor{myred!100}{\textbf{0.7080}} & \cellcolor{myred!50}{\underline{0.6442}} & 0.5945 & 0.4042 & 0.3203 & 0.4075 & 0.4790 & 0.6225 & 0.2240 \\
& Vanilla   &   \cellcolor{myred!100}{\textbf{0.5357}} & 0.5335 & 0.5177 & 0.3532 & 0.2757 & 0.5260 & 0.4311 & \cellcolor{myred!50}{\underline{0.5395}} & 0.2543 \\

\hline
\multicolumn{11}{c}{\cellcolor{color2}\it Design Repair} \\
\hline
\multirow{4}{*}{\bf MLLM Score} & React     & \cellcolor{myred!100}{\textbf{6.7857}} & 6.3571 & 6.3214 & 4.2143 & 2.7500 & \cellcolor{myred!50}{\underline{6.4643}} & 5.3571 & 5.6429 & 0.8929 \\
& Vue       & \cellcolor{myred!100}{\textbf{6.5926}} & 6.2593 & 6.0741 & 4.7778 & 3.5185 & \cellcolor{myred!50}{\underline{6.3704}} & 6.0370 & 6.0370 & 0.4815 \\
& Angular   & \cellcolor{myred!100}{\textbf{6.8571}} & 5.9286 & 5.2857 & 4.6429 & 3.2500 & \cellcolor{myred!50}{\underline{6.5357}} & 5.6429 & 6.5000 & 0.0000 \\
& Vanilla   & \cellcolor{myred!50}{\underline{7.1786}} & 7.0714 & \cellcolor{myred!100}{\textbf{7.3214}} & 5.7143 & 5.7857 & 6.9643 & 6.6786 & 6.8929 & 3.8571 \\
\hline
% \multirow{4}{*}{\bf CMS} & React     & \cellcolor{myred!100}{\textbf{0.4827}} & \cellcolor{myred!50}{\underline{0.2752}} & 0.1755 & 0.0448 & 0.0473 & 0.2272 & 0.0905 & 0.1866 & 0.0417 \\
% & Vue       & \cellcolor{myred!100}{\textbf{0.3065}} & \cellcolor{myred!50}{\underline{0.2524}} & 0.1782 & 0.0501 & 0.0474 & 0.2230 & 0.1557 & 0.1131 & 0.0127 \\
% & Angular   & \cellcolor{myred!100}{\textbf{0.5719}} & 0.5073 & 0.3968 & 0.3099 & 0.2688 & 0.2618 & 0.2546 & \cellcolor{myred!50}{\underline{0.5563}} & 0.0000 \\
% & Vanilla   & \cellcolor{myred!100}{\textbf{0.2287}} & \cellcolor{myred!50}{\underline{0.1637}} & 0.1630 & 0.0365 & 0.0690 & 0.1398 & 0.1188 & 0.1446 & 0.0277 \\ \hline

\multirow{3}{*}{\bf CSR} & React & \cellcolor{myred!100}{\textbf{1.0000}} & \cellcolor{myred!100}{\textbf{1.0000}} & \cellcolor{myred!100}{\textbf{1.0000}} & 0.9286 & 0.9643 & \cellcolor{myred!100}{\textbf{1.0000}} & 0.9643 & 0.9286 & 0.2857 \\
& Vue  & \cellcolor{myred!100}{\textbf{1.0000}} & \cellcolor{myred!100}{\textbf{1.0000}} & 0.9630 & \cellcolor{myred!100}{\textbf{1.0000}} & 0.8889 & \cellcolor{myred!100}{\textbf{1.0000}} & \cellcolor{myred!100}{\textbf{1.0000}} & \cellcolor{myred!100}{\textbf{1.0000}} & 0.1111 \\
& Angular  & 0.9286 & \cellcolor{myred!100}{\textbf{1.0000}} & \cellcolor{myred!100}{\textbf{1.0000}} & 0.7857 & 0.8571 & \cellcolor{myred!100}{\textbf{1.0000}} & \cellcolor{myred!100}{\textbf{1.0000}} & 0.9286 & 0.0357 \\ \hline

\multirow{4}{*}{\bf CMLS} & React     & \cellcolor{myred!100}{\textbf{0.6340}} & \cellcolor{myred!50}{\underline{0.4833}} & 0.3536 & 0.1389 & 0.0666  & 0.4551 & 0.1959  & 0.3616 & 0.0784 \\
& Vue       & \cellcolor{myred!100}{\textbf{0.5089}} & \underline{\cellcolor{myred!50}{0.4865}} & 0.3489 & 0.1533 & 0.1494 & 0.4422 &  0.3931  & 0.2451 & 0.0635 \\
& Angular   & \cellcolor{myred!100}{\textbf{0.6759}} & 0.5566 & 0.4785 & 0.3876 & 0.2729 & 0.5086 & 0.4995 & \cellcolor{myred!50}{\underline{0.6414}} & 0.0898 \\
& Vanilla   & \cellcolor{myred!50}{\underline{0.6088}} & \cellcolor{myred!100}{\textbf{0.6659}} & 0.4867 & 0.4237 & 0.2759 & 0.5563 & 0.5447 & 0.4990 & 0.1503 \\ \hline

\multirow{4}{*}{\bf CMCS} & React     & \cellcolor{myred!100}{\textbf{0.5420}} & \cellcolor{myred!50}{\underline{0.3872}} & 0.2548 & 0.0881 & 0.0365  & 0.3411 & 0.1331  & 0.2567 & 0.0563 \\
& Vue       & \cellcolor{myred!100}{\textbf{0.3924}} & \underline{ \cellcolor{myred!50}{0.3748}} & 0.2880  & 0.1024 & 0.0912 &  0.3377  & 0.3244 & 0.1800 & 0.0349 \\
& Angular   & \cellcolor{myred!50}{\underline{0.5706}} & 0.4527 & 0.3617 & 0.2711 & 0.1463 & 0.3894 & 0.3606 & \cellcolor{myred!100}{\textbf{0.5773}} & 0.0368 \\
& Vanilla   & \cellcolor{myred!50}{\underline{0.5859}} & \cellcolor{myred!100}{\textbf{0.6402}} & 0.4578 & 0.3780 & 0.2158 & 0.5404 & 0.5279 & 0.4780 & 0.1172 \\
\bottomrule
\end{tabular}}
\end{table*}

\subsection{RQ1:  How do MLLMs perform across distinct front-end tasks?}

% \begin{tcolorbox}[enhanced, colback=white, colframe=black, arc=10pt, boxrule=1pt]
% \textbf{Finding 1:} The best average of \textbf{91.73\%, 72.33\%, and 70.92\%} Completion@1, Compilation@1, and Pass@1, respectively, could be achieved on JavaBench over the studied LLMs. The Top-3 performing LLMs were DeepSeek-Coder-33b, gpt-3.5-turbo, and DeepSeek-Coder-6.7b among the studied LLMs.
% \end{tcolorbox}

Table~\ref{tab:overall} presents the performance of nine MLLMs on three tasks in three front-end frameworks React, Vue, and Angular. Vanilla denotes the webpage developed by vanilla HTML/CSS.

% For the Design Generation task, the bottleneck of the model is to understand the syntax and framework of the code developed in a framework-based manner.

% For the Design Edit and Design Repair tasks, it is difficult for the model to accurately locate the part that needs to be modified.

Among all, Claude-3.7, GPT-4o, Gemini-2.0, and Pixtral-124B are the top-performing MLLMs in three tasks.
Claude-3.7 achieves the highest performance across most metrics, including superior CLIP scores (0.6024-0.8319) for Design Generation and exceptional MLLM scores for Design Edit (8.01-9.15) and Design Repair (6.59-7.17). GPT-4o follows closely with strong CLIP scores (0.5963-0.7734) and outstanding compilation rates (0.7108-0.9725), alongside competitive MLLM scores across all tasks. Gemini-2.0 demonstrates solid performance with CLIP scores ranging from 0.6006-0.7611, reliable compilation success rates consistently above 0.71, and strong MLLM scores in Design Edit (7.81-9.13) and Design Repair (5.28-7.32). Pixtral-124B rounds out the top tier with competitive performance across multiple metrics, achieving strong CLIP scores (0.6324-0.7811), excellent compilation rates (0.7590-0.9746), and robust MLLM scores for Design Edit (8.01-9.11) and Design Repair (6.37-6.96).

\begin{tcolorbox}[colback=gray!20, colframe=gray!20, width=\columnwidth, left=0.05in, right=0.05in, top=0.05in, bottom=0.05in]
\textbf{Finding 1:} Among the evaluated MLLMs, Claude-3.7, GPT-4o, Gemini-2.0, and Pixtral-124B consistently demonstrated top-tier performance across the three tasks.
\end{tcolorbox}

% \begin{tcolorbox}[colback=gray!20, colframe=gray!20, width=\columnwidth]
% \textbf{Finding 1:} The best of xxxx, xxxx, and xxxx CLIP, Compilation Rate, and MLLM Score, respectively, could be achieved on \bench \ over the studied LLMs. The Top-3 performing MLLMs are claude-3-7-sonnet, gpt-4o, and gemini-2.0-flash among the studied MLLMs.
% \end{tcolorbox}

Larger variants consistently outperform their smaller counterparts within the same family. This is evident in comparisons such as Llama-90B versus Llama-11B, Pixtral-124B versus Pixtral-12B, and Qwen-72B versus Qwen-7B comparisons. The performance advantages are particularly pronounced in complex tasks requiring code localization and visual understanding, suggesting that increased model capacity enhances essential web development capabilities.

% \begin{tcolorbox}[colback=gray!20, colframe=gray!20, width=\columnwidth]
% \textbf{Finding 2:}
% Larger models consistently outperform smaller variants within the same family, demonstrating that increased model capacity significantly enhances web development capabilities.
% \end{tcolorbox}

\begin{tcolorbox}[colback=gray!20, colframe=gray!20, width=\columnwidth, left=0.05in, right=0.05in, top=0.05in, bottom=0.05in]
\textbf{Finding 2:}
Larger models consistently outperform smaller variants, demonstrating that increased model capacity enhances web development capabilities.
\end{tcolorbox}

In Design Generation tasks, MLLMs face two primary bottlenecks: compilation errors and visual inaccuracies. The compilation rates reveal significant framework-dependent challenges, with Angular showing the lowest success rates (0.6867-0.7590 for top models) compared to React and Vue ($>$0.83 for top models). Additionally, the moderate CLIP scores (0.5963-0.6324 even for best performers) indicate substantial room for improvement in generating visually accurate webpage layouts. The performance gap between vanilla HTML (highest CLIP scores) and framework-based implementations further suggests that the syntactic complexity in modern frameworks exacerbates challenges in both compilation and visual rendering.

In the Design Edit and Design Repair tasks, the primary bottleneck for MLLMs lies in accurately localizing and modifying the code segments that require changes. This limitation is reflected by the substantially lower CMLS and CMCS scores compared to compilation success rates across all models. In the Design Edit task, even top-performing models such as Claude-3.7 achieve CMLS scores of only 0.5555–0.7696 and CMCS scores of 0.4551–0.7080. A similar trend is observed in the Design Repair task, where Claude-3.7 attains CMLS values ranging from 0.5089 to 0.6759 and CMCS values from 0.3924 to 0.5859. These results indicate that, despite generating code that successfully compiles, MLLMs still struggle to precisely identify and semantically modify the target code regions.

% In Design Edit and Design Repair tasks, the primary bottleneck of MLLMs is the accurate localization and modification of code segments requiring modification. This is evidenced by CMLS and CMCS that are lower than compilation rates across all models. In Design Edit, even top-performing models like Claude-3.7 achieve CMLS of only 0.5555-0.7696 and CMCS of only 0.4551-0.7080. Similarly, in Design Repair, CMLS scores range from 0.4245-0.8071 for Claude-3.7 and CMCS range from 0.3537 to 0.7378. These results highlight a substantial difficulty in localizing target code, even when the generated code successfully compiles.

\begin{tcolorbox}[colback=gray!20, colframe=gray!20, width=\columnwidth, left=0.05in, right=0.05in, top=0.05in, bottom=0.05in]
\textbf{Finding 3:} MLLMs exhibit task-specific bottlenecks: Design Generation is challenged by compilation errors and visual inaccuracies, while Design Edit and Repair are mainly limited by deficiencies in code localization and modification.
% MLLMs face distinct bottlenecks across design tasks. In Design Generation, the primary challenges are eliminating compilation errors and achieving accurate visual rendering. For Design Edit and Repair tasks, the main limitation is accurately localizing the specific code segments that require modification.
\end{tcolorbox}

\subsection{RQ2: What is the comparative performance of MLLMs when applied to different development frameworks?}

\begin{figure}[t]
    \subfigure[CLIP.]{
    \label{fig:CLIP}
    \centering
    \includegraphics[width = .3\textwidth]{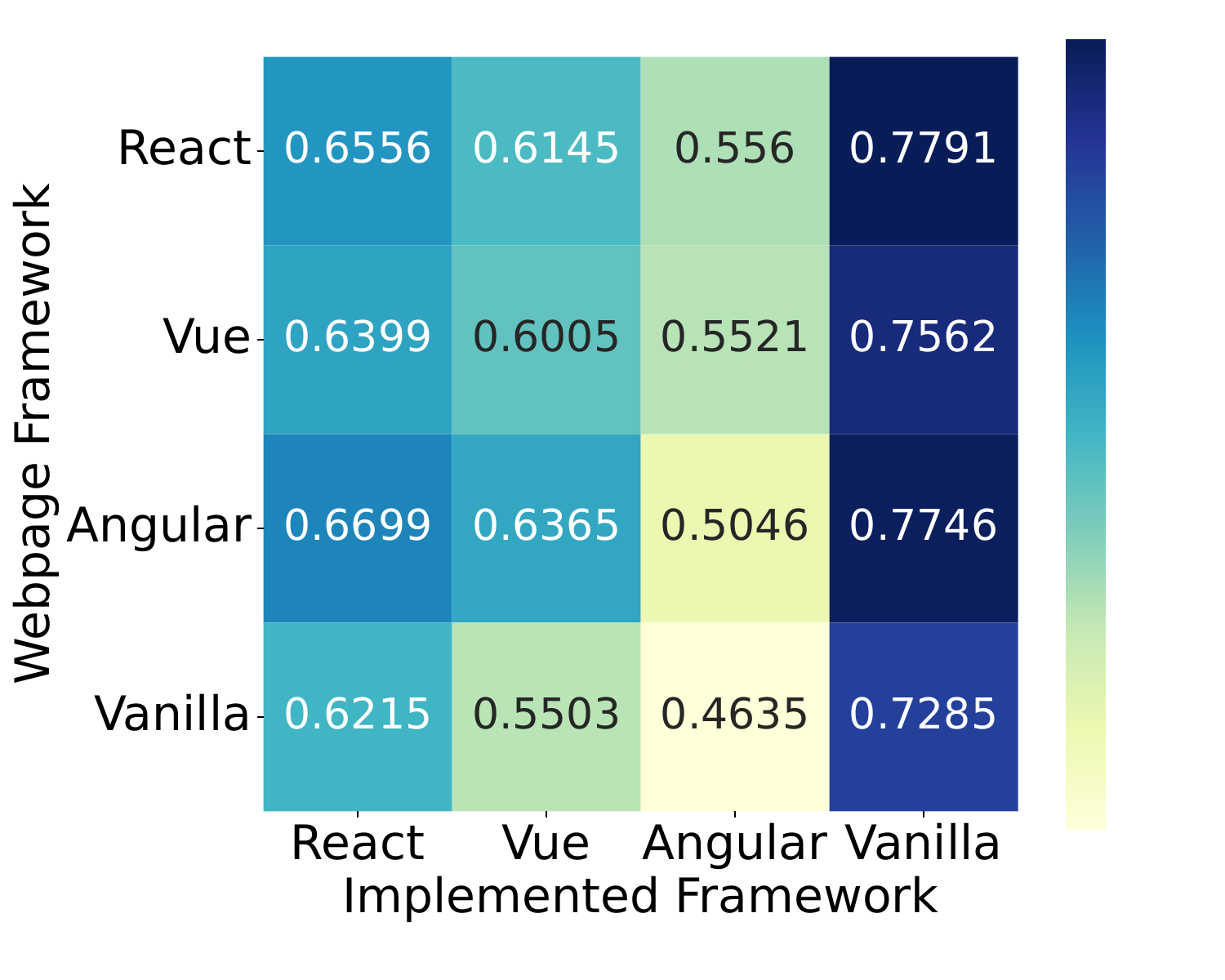}
    }
    \subfigure[Compilation.]{
    \label{fig:Compile}
    \centering
    \includegraphics[width = .3\textwidth]{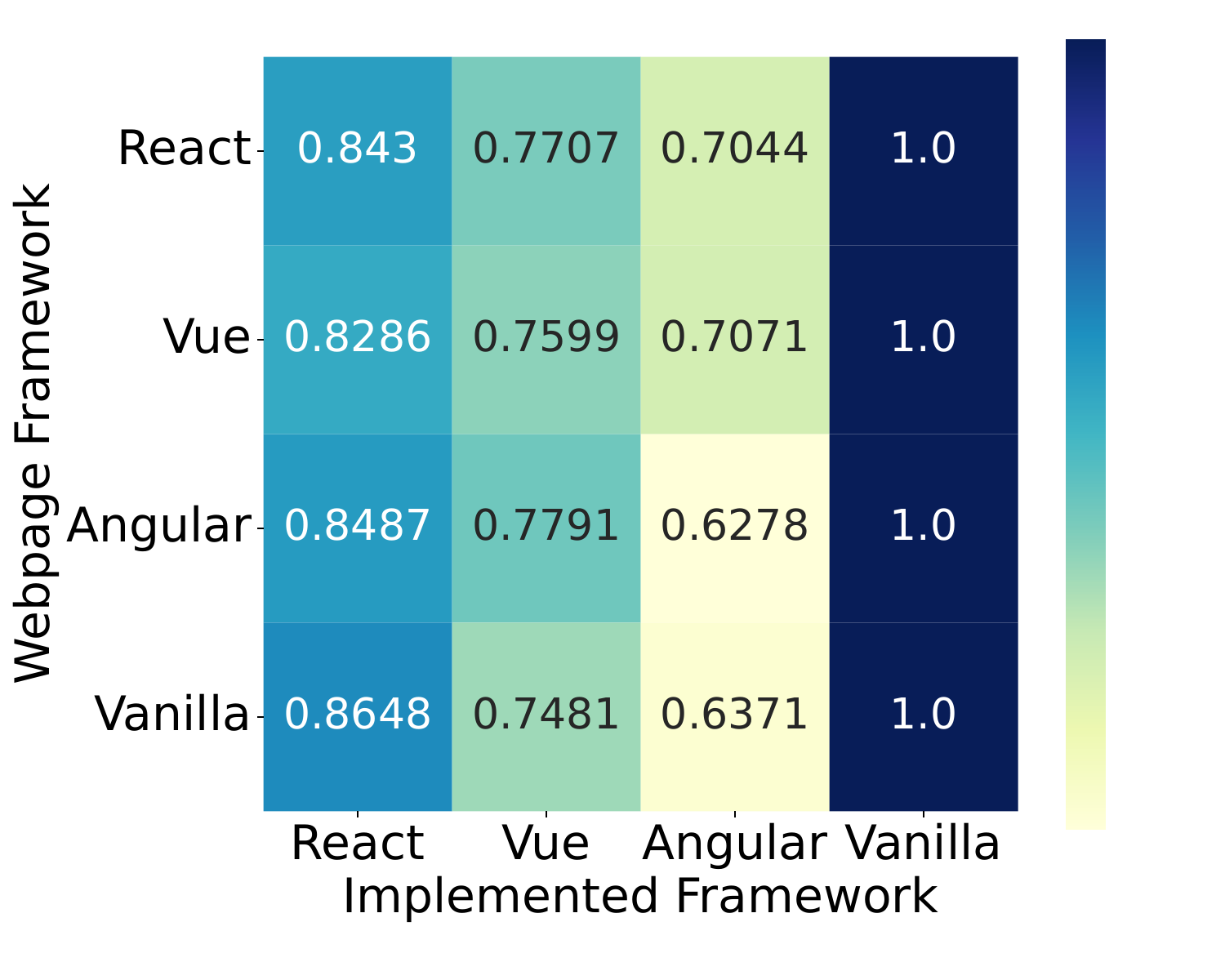}
    }
    \caption{Performance of webpages implemented by different frameworks. y-axis denotes the framework used to actually implement the webpage and x-axis represents the framework used by the model.}
    % \vspace{-0.1in}
    \label{fig:framework}
    \vspace{-0.1in}
\end{figure}

To further explore the model's proficiency across different front-end frameworks, we evaluate MLLMs' ability to implement webpages using various framework combinations.

Fig.~\ref{fig:framework} presents the average CLIP scores and compilation success rates of nine MLLMs across different framework combinations. The results reveal distinct performance patterns across frameworks. MLLMs consistently achieve optimal performance when implementing webpages using vanilla HTML/CSS, attaining the highest CLIP scores above 0.72 and perfect compilation success rates. In contrast, Angular-based implementations demonstrate the poorest performance, with compilation success rates ranging from 0.6-0.7 and CLIP scores between 0.45-0.55. React and Vue frameworks show intermediate performance levels, with both achieving reasonable compilation rates and moderate CLIP scores, though still inferior to vanilla implementations.

\begin{tcolorbox}[colback=gray!20, colframe=gray!20, width=\columnwidth, left=0.05in, right=0.05in, top=0.05in, bottom=0.05in]
\textbf{Finding 4:}
MLLMs demonstrate the strongest performance with vanilla HTML/CSS, followed by React and Vue, but exhibit significant challenges with Angular implementations.
% MLLMs demonstrate superior performance in generating native HTML/CSS code compared to framework-based implementations. Among the three frameworks evaluated, MLLMs exhibit strong proficiency with React and Vue, but show significantly weaker performance with Angular framework.
\end{tcolorbox}

\subsection{RQ3: How does varying task difficulty impact MLLM performance?}

\begin{table}[h]
\centering
\footnotesize
\caption{Performance under different difficulty levels. }
\setlength{\tabcolsep}{0.2em}
\label{tab:difficulty}
\begin{tabular}{c|ccc|ccc|ccc}
\toprule
\multirow{2}{*}{\bf Model} & \multicolumn{3}{c|}{\it Design Generation} & \multicolumn{3}{c|}{\it Design Edit} & \multicolumn{3}{c}{\it Design Repair} \\
\cmidrule(lr){2-4} \cmidrule(lr){5-7} \cmidrule(lr){8-10}
& Easy & Medium & Hard & Easy & Medium & Hard & Easy & Medium & Hard \\
\midrule
Claude-3.7             & \cellcolor{myred!100}{\textbf{0.86}}      & \cellcolor{myred!70}{\underline{0.77}}     & \cellcolor{myred!25}{0.51}     & \cellcolor{myred!100}{\textbf{8.64}}                    & \cellcolor{myred!70}{\underline{8.32}}                     & \cellcolor{myred!25}{7.61}                   & \cellcolor{myred!100}{\textbf{7.21}} & \cellcolor{myred!25}{6.81} & \cellcolor{myred!70}{\underline{6.93}} \\
GPT-4o                 & \cellcolor{myred!100}{\textbf{0.84}}      & \cellcolor{myred!70}{\underline{0.74}}     & \cellcolor{myred!25}{0.43}     & \cellcolor{myred!100}{\textbf{8.73}}                   & \cellcolor{myred!25}{7.95}                     & \cellcolor{myred!70}{\underline{8.19}}                   & \cellcolor{myred!100}{\textbf{6.93}} & \cellcolor{myred!70}{\underline{6.46}} & \cellcolor{myred!25}{5.53} \\
Gemini-2.0             & \cellcolor{myred!100}{\textbf{0.84}}      & \cellcolor{myred!70}{\underline{0.74}}     & \cellcolor{myred!25}{0.743}     & \cellcolor{myred!100}{\textbf{9.05}}                   & \cellcolor{myred!70}{\underline{7.84}}                     & \cellcolor{myred!25}{7.51}                   & \cellcolor{myred!100}{\textbf{7.07}} & \cellcolor{myred!70}{\underline{6.21}} & \cellcolor{myred!25}{4.19} \\
Llama-90B              & \cellcolor{myred!100}{\textbf{0.69}}      & \cellcolor{myred!70}{\underline{0.65}}     & \cellcolor{myred!25}{0.47}     & \cellcolor{myred!70}{\underline{6.54}}                   & \cellcolor{myred!100}{\textbf{6.83}}                     & \cellcolor{myred!25}{5.68}                    & \cellcolor{myred!100}{\textbf{5.35}} & \cellcolor{myred!70}{\underline{4.90}} & \cellcolor{myred!25}{4.39} \\
Llama-11B              & \cellcolor{myred!100}{\textbf{0.68}}      & \cellcolor{myred!70}{\underline{0.62}}      & \cellcolor{myred!25}{0.46}     & \cellcolor{myred!100}{\textbf{5.88}}                    & \cellcolor{myred!70}{\underline{5.05}}                     & \cellcolor{myred!25}{5.24}                   & \cellcolor{myred!70}{\underline{3.91}} & \cellcolor{myred!100}{\textbf{4.07}} & \cellcolor{myred!25}{2.41} \\
Pixtral-124B           & \cellcolor{myred!100}{\textbf{0.78}}      & \cellcolor{myred!70}{\underline{0.73}}     & \cellcolor{myred!25}{0.53}     & \cellcolor{myred!100}{\textbf{8.69}}                   & \cellcolor{myred!70}{\underline{7.91}}                     & \cellcolor{myred!25}{6.98}                   & \cellcolor{myred!100}{\textbf{7.30}} & \cellcolor{myred!70}{\underline{6.56}} & \cellcolor{myred!25}{5.66} \\
Pixtral-12B            & \cellcolor{myred!70}{\underline{0.67}}      & \cellcolor{myred!100}{\textbf{0.68}}     & \cellcolor{myred!25}{0.50}     & \cellcolor{myred!100}{\textbf{8.26}}                   & \cellcolor{myred!70}{\underline{7.63}}                     & \cellcolor{myred!25}{6.71}                   & \cellcolor{myred!100}{\textbf{6.76}} & \cellcolor{myred!70}{\underline{5.84}} & \cellcolor{myred!25}{5.30} \\
Qwen-72B               & \cellcolor{myred!100}{\textbf{0.83}}      & \cellcolor{myred!70}{\underline{0.71}}     & \cellcolor{myred!25}{0.44}     & \cellcolor{myred!100}{\textbf{7.78}}                   & \cellcolor{myred!70}{\underline{6.81}}                     & \cellcolor{myred!25}{6.72}                   & \cellcolor{myred!100}{\textbf{7.12}} & \cellcolor{myred!70}{\underline{6.06}} & \cellcolor{myred!25}{5.57} \\
Qwen-7B                & \cellcolor{myred!100}{\textbf{0.56}}      & \cellcolor{myred!70}{\underline{0.53}}     & \cellcolor{myred!25}{0.32}     & \cellcolor{myred!100}{\textbf{3.46}}                   & \cellcolor{myred!70}{\underline{2.83}}                     & \cellcolor{myred!25}{2.71}                   & \cellcolor{myred!70}{\underline{1.09}} & \cellcolor{myred!100}{\textbf{1.63}} & \cellcolor{myred!25}{1.05} \\
\bottomrule
\end{tabular}
\end{table}

We categorize samples into different difficulty levels to systematically evaluate MLLMs' performance across varying complexity scenarios. 

% The difficulty assessment criteria differ for each task to reflect their distinct challenges.

% For Design Generation task, which requires generating code from scratch, difficulty is primarily determined by the visual complexity of webpages. We classify webpages smaller than 1000$\times$1000 pixels as easy, those larger than 2000$\times$2000 pixels as hard, and intermediate sizes as medium difficulty.

For Design Generation task, which requires generating code from scratch, we calculate the complexity based on the image size $I$, UI elements number $U$, color variety $C$ (i.e., the number of colors in the UI) and layout complexity $L$ (i.e., number of layout methods + number of nesting levels + number of CSS layout attributes). We then normalize these scores to between 0 and 100. The complexity score: $S = 0.25 \times I + 0.25 \times U + 0.25 \times C + 0.25 \times L$. Scores greater than 80 are complex, less than 30 are simple, and between 30 and 80 are medium. 

In Design Edit task, where models must modify existing code according to user instructions, difficulty correlates with the complexity and scope of the modification instructions. We adopt the annotator-provided difficulty labels from Section~\ref{subsec:da} as our ground truth difficulty classification.

For Design Repair task, difficulty is assessed based on the severity of UI issues, quantified by the extent of code modifications required to resolve the identified problems. Tasks requiring modifications to more than 30 lines of code are classified as hard, those requiring fewer than 10 lines as easy, and intermediate cases as medium difficulty.

% We divide different samples into different levels of difficulty to measure the performance of MLLM on samples of different difficulty levels.

% However, the difficulty measurement indicators of the three tasks are different.

% For Design Generation task, which require generating code from scratch, difficulty is primarily measured by the visual complexity of the target webpage. Web pages smaller than 1000*1000 are considered simple, web pages larger than 2000*2000 are considered complex, and web pages in between are medium.

% In Design Edit task, where models must modify existing code according to user instructions, difficulty correlates with the complexity of users' instructions. 
% We directly take the label of Annotator in Section x as the difficulty label.

% For Design Repair tasks, difficulty is assessed based on the severity of UI issues, specifically quantified by the extent of code modifications necessary to resolve the identified problems. A number of lines greater than 30 that need to be modified is considered difficult, less than 10 is easy, and anything in between is moderate.

Table~\ref{tab:difficulty} shows the CLIP score for Design Generation task and MLLM score for Design Edit and Design Repair task under different difficulty levels.

Table~\ref{tab:difficulty} reveals distinct difficulty-related performance patterns across the three design tasks. In Design Generation, where difficulty stems from image complexity and size, top models show moderate degradation from ~0.79-0.83 (Easy) to ~0.69-0.73 (Hard). Design Edit tasks, with difficulty determined by the number of required operations, exhibit more pronounced drops, particularly for complex multi-operation scenarios where top models decline from 8.64-9.05 to 7.51-8.19. Design Repair shows the most severe degradation, where UI issue severity and required code modifications cause performance to plummet from 6.93-7.21 (Easy) to 4.19-6.93 (Hard), with smaller models experiencing catastrophic failures. These patterns indicate that visual complexity moderately affects generation, operational complexity significantly impacts editing, while code-level debugging presents the steepest performance barriers for current MLLMs.

\begin{tcolorbox}[colback=gray!20, colframe=gray!20, width=\columnwidth, left=0.05in, right=0.05in, top=0.05in, bottom=0.05in]
\textbf{Finding 5:}
MLLM performance degrades when confronted with large UI images in Design Generation, complex instructions in Design Edit, and severe UI issues in Design Repair tasks.
\end{tcolorbox}

\subsection{RQ4: To what extent do different input contexts affect MLLMs’ performance?}

For the Design Edit and Design Repair tasks, we explore the impact of input images and codes on the results. Table~\ref{tab:context} shows the MLLM score on in Design Edit and Design Repair tasks under different context inputs. The results reveal distinct patterns regarding the utility of visual versus code information across different models and tasks.

Code-only input consistently outperforms image-only input across both tasks and all models. For Design Edit, top models (Claude-3.7, GPT-4o, Gemini-2.0) achieve highest scores with code-only input (8.40-8.43) versus image-only (7.37-7.67). Design Repair shows similar code-only superiority (6.53-6.70 vs 5.47-5.81 for top models).

However, combining code and image inputs does not yield significant improvements and occasionally results in minor performance degradation, highlighting the limitations of MLLMs in accurately localizing modification points and identifying UI issues through visual analysis.

\begin{table}[t]
\centering
\footnotesize
\caption{Performance under different context inputs. Both denote combining image and code.}
\label{tab:context}
\setlength{\tabcolsep}{0.6em}
\begin{tabular}{c|ccc|ccc}
\toprule
\multirow{2}{*}{\bf Model} & \multicolumn{3}{c|}{\it Design Edit} & \multicolumn{3}{c}{\it Design Repair}  \\
\cmidrule(lr){2-4} \cmidrule(lr){5-7}
& Image & Code & Both & Image & Code & Both \\
\midrule
Claude-3.7             & 7.6764 & \textbf{8.4326} & 8.4258     & 5.8142 & 6.7014 & \textbf{6.8535} \\
GPT-4o                 & 7.3728 & 8.4013 & \textbf{8.4258}     & 5.6968 & \textbf{6.5304} & 6.4041 \\
Gemini-2.0             & 7.6430 & 8.4105 & \textbf{8.5083}     & 5.4712 & \textbf{6.6726} & 6.2506 \\
Llama-90B              & 4.9646 & \textbf{7.4939} & 6.4415     & 3.7860 & \textbf{5.3178} & 4.8373 \\
Llama-11B              & 2.5431 & \textbf{6.3324} & 4.9111     & 3.6713 & \textbf{4.9005} & 3.8260 \\
Pixtral-124B           & 7.6418 & \textbf{8.5663} & 8.4627     & 5.3816 & 6.5486 & \textbf{6.5837} \\
Pixtral-12B            & 6.2005 & 7.6840 & \textbf{7.7583}     & 4.8128 & \textbf{6.2159} & 5.9289 \\
Qwen-72B               & 4.7724 & 8.2313 & \textbf{8.2442}     & 5.3158 & \textbf{6.2308} & 6.2682 \\
Qwen-7B                & 1.8098 & 2.9874 & \textbf{ 3.0256}     & 1.4672 & \textbf{1.5225} & 1.3079 \\
\bottomrule
\end{tabular}
\end{table}

\begin{tcolorbox}[colback=gray!20, colframe=gray!20, width=\columnwidth, left=0.05in, right=0.05in, top=0.05in, bottom=0.05in]
\textbf{Finding 6:} Code-only input consistently outperforms image-only input, but combining both input types does not yield much improvement. This suggests that textual code offers MLLMs with more semantic information than visual data in Design edit and Repair task.
\end{tcolorbox}

% \begin{tcolorbox}[colback=gray!20, colframe=gray!20, width=\columnwidth]
% \textbf{Finding 6:}
% Rich contextual information significantly enhances the performance of MLLMs in design edit and desin repair tasks by combining code and image information. 
% \end{tcolorbox}

% \subsection{RQ5: What are the limitations of MLLMs when developing framework-based webpages?}

\subsection{RQ5: The limitations of MLLMs when developing framework-based webpages?}

\subsubsection{Compilation Errors} 
\begin{figure*}[ht]
    \subfigure[React.]{
    \label{fig:reacr}
    \centering
    \includegraphics[width = .3\textwidth]{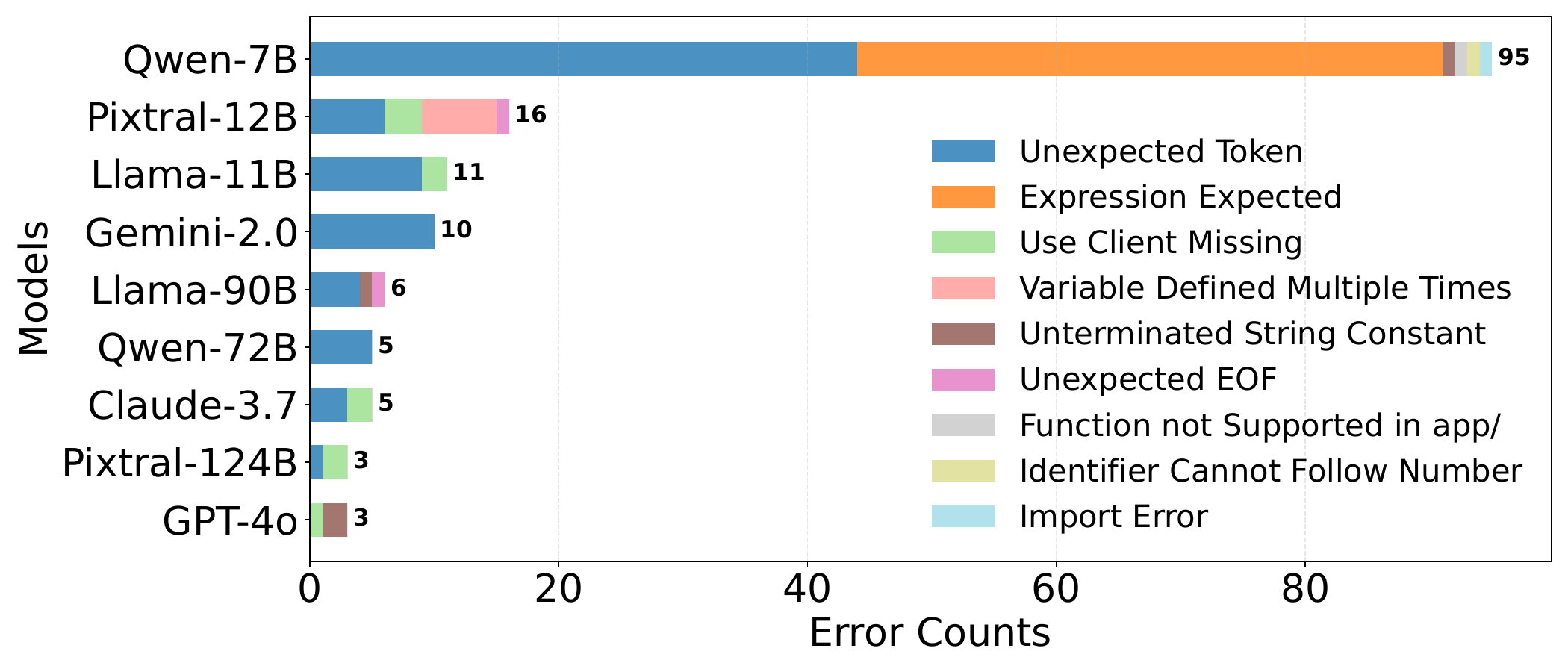}
    }
    \subfigure[Vue.]{
    \label{fig:vue}
    \centering
    \includegraphics[width = .3\textwidth]{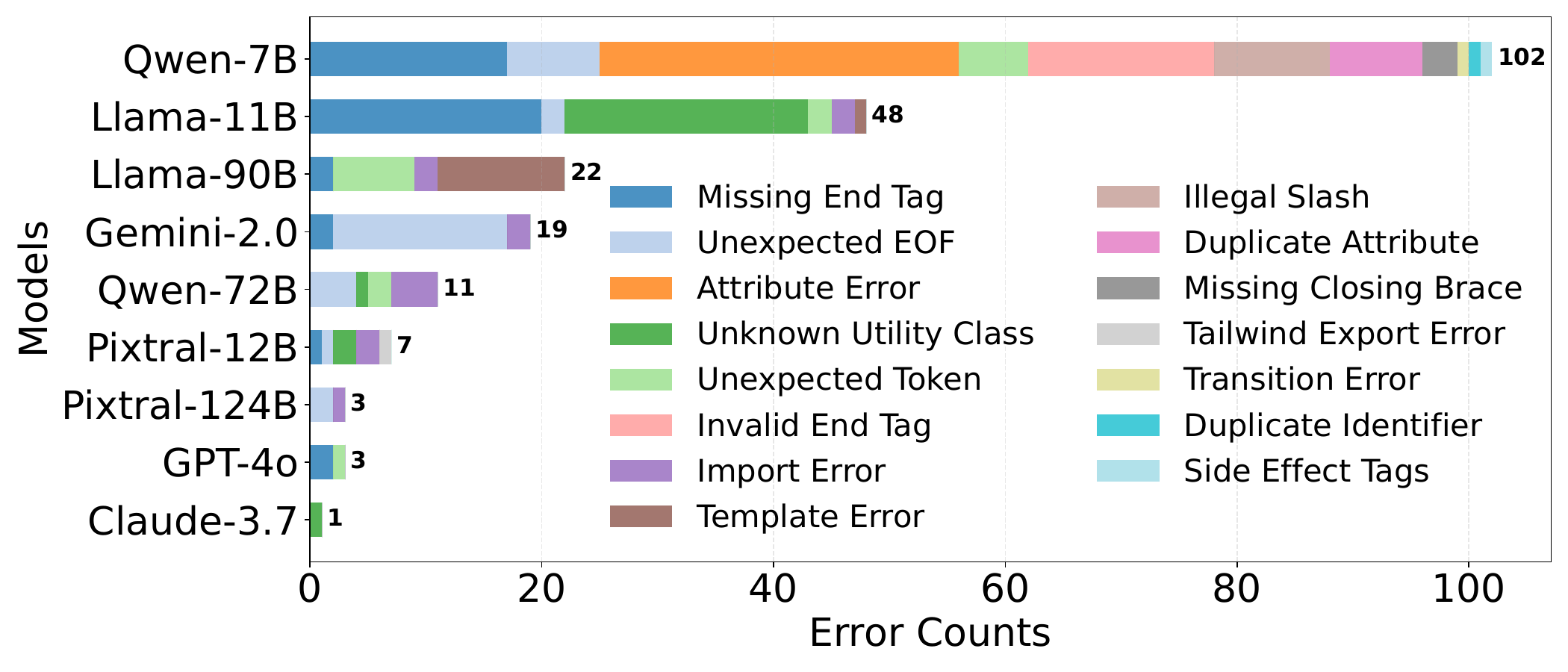}
    }
    \subfigure[Angular.]{
    \label{fig:angular}
    \centering
    \includegraphics[width = .3\textwidth]{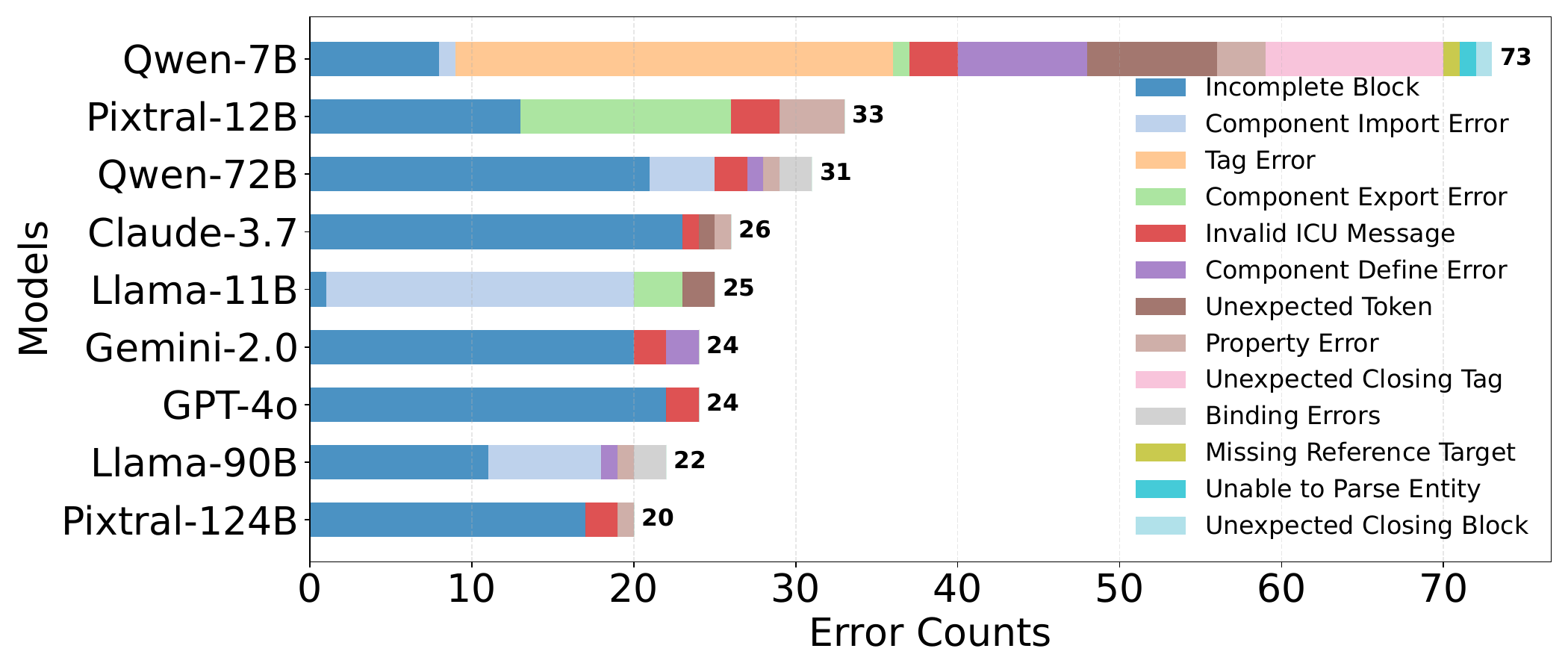}
    }
    \caption{Compilation error distribution.}
    \label{fig:errors}
\end{figure*}

The compilation error indicates that MLLM does not sufficiently understand the syntax of framework-based front-end development. As illustrated in Figure~\ref{fig:errors}, different MLLMs exhibit distinct error patterns across frameworks, revealing specific weaknesses in their understanding of front-end development syntax.

In React development, the top three errors are ``Unexpected Token'', ``Expression Expected'', and ``Use Client Missing'', with Qwen-7B producing the highest number of compilation errors (95 total) while GPT-4o demonstrates superior performance with only 3 total errors. For Vue development, ``Missing End Tag'', ``Unexpected EOF'', and ``Attribute Error'' are dominated errors. Qwen-7B shows the most compilation issues (102), while Claude-3.7 achieves near-perfect compilation success with just 1 error. Angular development presents a different error profile, with ``Incomplete Block'', ``Component Import Error'', and ``Component Export Error'' as the primary issues. Qwen-7B continues to struggle with 73 total errors, while Pixtral-124B shows the best performance.

% improved performance compared to other frameworks.

% The error distribution reveals framework-specific challenges: React primarily faces JSX parsing issues, Vue encounters template syntax problems, while Angular suffers from TypeScript-related component management difficulties. Advanced models like GPT-4o, Claude-3.7, and Pixtral-124B consistently demonstrate superior syntax understanding across all frameworks.

The error distribution reveals that MLLMs face distinct limitations when working with different frameworks: they struggle with JSX syntax parsing and React-specific expressions in React applications, encounter difficulties with template structure and attribute handling in Vue development, and show inadequate understanding of TypeScript module systems and component architecture in Angular projects. Advanced models like GPT-4o, Claude-3.7, and Pixtral-124B consistently demonstrate superior syntax understanding.

% suggesting that model scale and training sophistication significantly impact framework-specific code generation capabilities.

\begin{tcolorbox}[colback=gray!20, colframe=gray!20, width=\columnwidth, left=0.05in, right=0.05in, top=0.05in, bottom=0.05in]
\textbf{Finding 7:} MLLMs exhibit framework-specific limitations: struggling with JSX parsing in React, template syntax in Vue, and TypeScript components in Angular. Advanced models demonstrate significantly better syntax comprehension across all frameworks.
\end{tcolorbox}

To verify MLLM's ability to solve compilation errors, we sample 30 webpages with diverse compilation errors and prompt MLLMs to fix them. The results are shown in the Table~\ref{tab:compile}. The overall average repair rate across all models and frameworks is 0.53, indicating that MLLMs still face challenges in fixing front-end errors.

% \begin{table}[ht]
% \centering
% \caption{Compile error repair rate of MLLMs.}
% \label{tab:compile}
% \setlength{\tabcolsep}{0.1em}
% \begin{tabular}{c|cccc}
% \toprule
% \textbf{Model} & \textbf{React} & \textbf{Vue} & \textbf{Angular} & \textbf{Average} \\
% \midrule
% Claude-3.7 & 0.70 & 0.40 & 0.70 & 0.60 \\
% GPT-4o & 0.60 & 0.50 & 0.50 & 0.53 \\
% Gemini-2.0 & 0.60 & 0.80 & 0.80 & 0.73 \\
% Llama-90B & 0.50 & 0.60 & 0.80 & 0.63 \\
% Llama-11B & 0.30 & 0.50 & 0.20 & 0.33 \\
% Pixtral-124B & 0.70 & 0.70 & 0.70 & 0.70 \\
% Pixtral-12B & 0.50 & 0.60 & 0.40 & 0.50 \\
% Qwen-72B & 0.70 & 0.50 & 0.50 & 0.57 \\
% Qwen-7B & 0.20 & 0.10 & 0.20 & 0.17 \\
% \midrule
% \textbf{Average} & 0.53 & 0.52 & 0.53 & 0.53 \\
% \bottomrule
% \end{tabular}
% \end{table}

\begin{table}[t]
\centering
\footnotesize
\begin{minipage}{0.4\textwidth}
\centering
\caption{Compile error fix ratio.}
\label{tab:compile}
\setlength{\tabcolsep}{0.1em}
\begin{tabular}{c|cccc}
\toprule
\textbf{Model} & \textbf{React} & \textbf{Vue} & \textbf{Angular} \\
\midrule
Claude-3.7 & 0.70 & 0.40 & 0.70\\
GPT-4o & 0.60 & 0.50 & 0.50 \\
Gemini-2.0 & 0.60 & 0.80 & 0.80 \\
Llama-90B & 0.50 & 0.60 & 0.80 \\
Llama-11B & 0.30 & 0.50 & 0.20 \\
Pixtral-124B & 0.70 & 0.70 & 0.70 \\
Pixtral-12B & 0.50 & 0.60 & 0.40 \\
Qwen-72B & 0.70 & 0.50 & 0.50 \\
Qwen-7B & 0.20 & 0.10 & 0.20 \\
\midrule
\textbf{Average} & 0.53 & 0.52 & 0.53 \\
\bottomrule
\end{tabular}
\end{minipage}
% \hfill
% \hspace{0.1em}
\begin{minipage}{0.52\textwidth}
\centering
% \caption{Percentage of webpages implemented by MLLMs using component-based design.}
\caption{Ratio of using component-based design.}
\label{tab:component}
\setlength{\tabcolsep}{0.1em}
\begin{tabular}{c|cccc}
\toprule
\textbf{Model} & \textbf{React} & \textbf{Vue} & \textbf{Angular} \\
\midrule
Claude-3.7 & 0.23\% & 6\% & 38\% \\
GPT-4o & \textbf{0.71\%} & \textbf{6.3\%} & 10\% \\
Gemini-2.0 & 0.7\% & 0.23\% & \textbf{41\%} \\
Llama-90B & 0\% & 2\% & 5\% \\
Llama-11B & 0.48\% & 0.47\% & 1.3\% \\
Pixtral-124B & 0\% & 5.3\% & 7.7\% \\
Pixtral-12B & 0\% & 1.4\% & 2.3\% \\
Qwen-72B & 0\% & 17\% & 28\% \\
Qwen-7B & 0\% & 6\% & 40\% \\
\hline
Average & 0.24\% & 5\% & 19\% \\
\bottomrule
\end{tabular}
\end{minipage}
\end{table}

% \begin{table}[]
% \centering
% \caption{Compile error repair rate of MLLMs.}
% \label{tab:compile}
% \begin{tabular}{c|cccc}
% \toprule
% \textbf{Model} & \textbf{React} & \textbf{Vue} & \textbf{Angular} \\
% \midrule
% Claude-3.7 & 7/10 & 4/10 & 7/10 \\
% GPT-4o & 6/10 & 5/10 & 5/10 \\
% Gemini-2.0 & 6/10 & 8/10 & 8/10 \\
% Llama-90B & 5/10 & 6/10 & 8/10 \\
% Llama-11B & 3/10 & 5/10 & 2/10 \\
% Pixtral-124B & 7/10 & 7/10 & 7/10 \\
% Pixtral-12B & 5/10 & 6/10 & 4/10 \\
% Qwen-72B & 7/10 & 5/10 & 5/10 \\
% Qwen-7B & 2/10 & 1/10 & 2/10 \\
% % \hline
% % Average & 0.24\% & 5\% & 19\% \\
% \bottomrule
% \end{tabular}
% \end{table}

\subsubsection{Component-based Implementation Limitation} Component-based design is a method of breaking down user interfaces into reusable, self-contained parts called components. This approach improves efficiency and scalability in website development. Table~\ref{tab:component} shows the proportion of webpages implemented by MLLMs using component-based design across three frameworks. The result reveals that MLLMs demonstrate remarkably low adoption rates of component-based implementation, with average 0.24\%, 5\% and 19\% rate on React, Vue and Angular, respectively.

% \begin{table}[ht]
% \centering
% \caption{Percentage of webpages implemented by MLLMs using component-based design.}
% \label{tab:component}
% \setlength{\tabcolsep}{0.1em}
% \begin{tabular}{c|cccc}
% \toprule
% \textbf{Model} & \textbf{React} & \textbf{Vue} & \textbf{Angular} \\
% \midrule
% Claude-3.7 & 0.23\% & 6\% & 38\% \\
% GPT-4o & \textbf{0.71\%} & \textbf{6.3\%} & 10\% \\
% Gemini-2.0 & 0.7\% & 0.23\% & \textbf{41\%} \\
% Llama-90B & 0\% & 2\% & 5\% \\
% Llama-11B & 0.48\% & 0.47\% & 1.3\% \\
% Pixtral-124B & 0\% & 5.3\% & 7.7\% \\
% Pixtral-12B & 0\% & 1.4\% & 2.3\% \\
% Qwen-72B & 0\% & 17\% & 28\% \\
% Qwen-7B & 0\% & 6\% & 40\% \\
% \hline
% Average & 0.24\% & 5\% & 19\% \\
% \bottomrule
% \end{tabular}
% \end{table}

% \textbf{Case Study.} Fig.~\ref{fig:case} shows a screenshot of the Recycle Live section on the BBC website. It is obvious that each item has the same structure and style, so the best approach is to implement a news component and then reuse it in a loop manner.
% However, as shown in Listing~\ref{listing:vue}, the Vue code generated by MLLMs contains repeated items, which shows that MLLM’s understanding of component-based implementation and framework-specific syntax (such as v-for) is insufficient.

\begin{figure}[ht]
\centering
\begin{minipage}{0.45\textwidth}
\centering
\includegraphics[width=\textwidth]{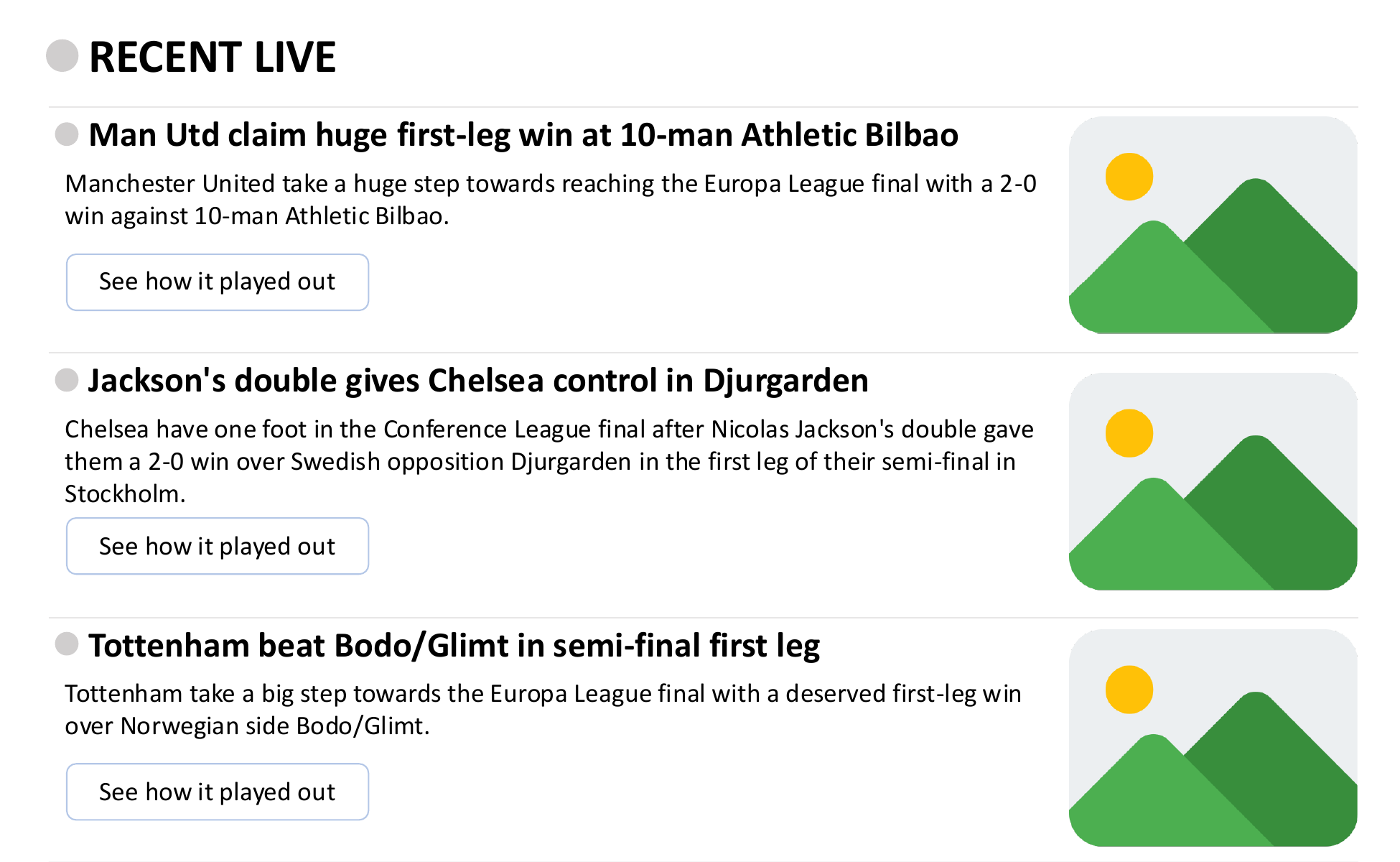}
\caption{A webpage with multiple elements of the same structure and style.}
\label{fig:case}
\end{minipage}%
\hfill
\begin{minipage}{0.5\textwidth}
\definecolor{stubbg}{HTML}{FCF3D5}
\begin{lstlisting}[
   language=html,
   literate={\ }{{\phantom{x}}}1,
   aboveskip=-2pt,
   belowskip=-10pt,
   caption={Vue implementation containing repeated items.},
   label={listing:vue},
   escapechar=|,
   numbers=left
]
<!-- Item 1 -->
<div class="border-l-2 border-gray-300 pl-4">
  <div class="flex">
    <div class="flex-1 pr-4">
      <h3 class="font-bold text-lg mb-1">Man Utd claim huge first-leg win...</h3>
      <p class="text-sm mb-2">Manchester United take a huge step towards...</p>
      <button class="text-xs border border-gray-300 px-2 py-1 hover:bg-gray-100"> See how it played out </button>
    </div>
    <div class="w-1/3">
      <img src="./place.svg" alt="Man Utd vs Athletic Bilbao" class="w-full" />
    </div>
  </div>
</div>
<!-- Item 2 -->
<!-- Additional items... -->
\end{lstlisting}
\end{minipage}
\end{figure}

\textbf{Case Study.} Fig.~\ref{fig:case} shows a screenshot of the BBC website's recently live section. Each news item follows an identical structure, making it ideal for component-based implementation with iterative rendering. However, as shown in Listing~\ref{listing:vue}, the Vue code generated by MLLMs contains hardcoded, repetitive structures instead of utilizing Vue's \texttt{v-for} directive. This reveals MLLMs' insufficient understanding of component-based architecture and framework-specific syntax.

% \begin{figure}[ht]
% \centering
% \includegraphics[width = .4\textwidth]{figures/case1.pdf}
% \caption{A webpage with multiple elements of the same structure and style.}
% \label{fig:case}
% \end{figure}

% \definecolor{stubbg}{HTML}{FCF3D5}

% \begin{figure}[h!]
% 	\lstinputlisting[
% 		language=html,
%         literate={\ }{{\phantom{x}}}1,
%         % basicstyle=\linespread{0.8}\tiny,
%         % baselinestretch=1.0,
% 		% morekeywords={var},
%             aboveskip=-2pt,
%             belowskip=-10pt,
% 		caption={Vue implementation containing repeated items.},
% 		label={listing:vue},
% 		escapechar=|,
% 		% linebackgroundcolor = {
%   %               \ifnum \value{lstnumber} = 12 \color{removed} \fi
%   %               \ifnum \value{lstnumber} = 13 \color{added} \fi
%   %           },
% 		numbers=left
% 	]{./code/vue.tex}
% \end{figure}

% \begin{figure}[h!]
% 	\lstinputlisting[
% 		language=html,
%         % basicstyle=\linespread{0.8}\tiny,
%         % baselinestretch=1.0,
% 		% morekeywords={var},
%             % aboveskip=-2pt,
%             % belowskip=-10pt,
% 		caption={Angular Implementation},
% 		label={listing:ce2},
% 		% escapechar=|,
% 		% linebackgroundcolor = {
%   %               \ifnum \value{lstnumber} = 12 \color{removed} \fi
%   %               \ifnum \value{lstnumber} = 13 \color{added} \fi
%   %           },
% 		numbers=left
% 	]{./code/angular.tex}
% \end{figure}

% \begin{tcolorbox}[colback=gray!20, colframe=gray!20, width=\columnwidth]
% \textbf{Finding 8:}
% MLLMs exhibit a critical deficiency in generating component-based implementations, highlighting the models' limitation to produce reuable and maintainable front-end code.
% \end{tcolorbox}

\begin{tcolorbox}[colback=gray!20, colframe=gray!20, width=\columnwidth, left=0.05in, right=0.05in, top=0.05in, bottom=0.05in]
\textbf{Finding 8:}
MLLMs show critical deficiencies in component-based implementation and framework-specific syntax, revealing limitations in producing reusable front-end code.
\end{tcolorbox}
% that leverages each framework's distinctive capabilities.

\subsubsection{Issue Detection Limitation} 

In Design Repair task, we also prompt MLLMs to judge the UI display issues described in Section \ref{subsec:da}. Then we calculate the issue detection accuracy by comparing the MLLMs' outputs with the ground truth annotated by human experts. Table~\ref{tab:mllm_issue_detection} presents the UI issue detection accuracy across different MLLM models. The results demonstrate consistently poor performance of MLLMs in identifying UI issues, with average 0.2972, 0.2205 and 0.2275, 0.3403 rate on React, Vue, Angular and Vanilla, respectively.

\begin{table}[t]
\centering
\footnotesize
\caption{UI Issue Identification Accuracy of different MLLMs.}
\label{tab:mllm_issue_detection}
\begin{tabular}{l|rrrr|r}
\toprule
\textbf{Model} & \textbf{React} & \textbf{Vue} & \textbf{Angular} & \textbf{Vanilla} & \textbf{Average} \\
\midrule
Claude-3.7 & 0.4155 & 0.2654 & 0.3929 & 0.4286 & 0.3756 \\
GPT-4o & \textbf{0.4369} & 0.2870 & \textbf{0.4101} & 0.4464 & \textbf{0.3951} \\
Gemini-2.0 & 0.4250 & 0.3210 & 0.2649 & \textbf{0.5417} & 0.3882 \\
Llama-90B & 0.2827 & \textbf{0.3735} & 0.1470 & 0.3571 & 0.2901 \\
Llama-11B & 0.0179 & 0.0000 & 0.0298 & 0.0119 & 0.0149 \\
Pixtral-124B & 0.3315 & 0.2747 & 0.3488 & 0.3988 & 0.3385 \\
Pixtral-12B & 0.3405 & 0.2099 & 0.0685 & 0.2804 & 0.2248 \\
Qwen-72b & 0.3881 & 0.2222 & 0.3851 & 0.3631 & 0.3396 \\
Qwen-7b & 0.0369 & 0.0309 & 0.0000 & 0.2351 & 0.0757 \\
\midrule
\textbf{Average} & 0.2972 & 0.2205 & 0.2275 & 0.3403 & 0.2714 \\
\bottomrule
\end{tabular}
% \vspace{-0.1in}
\end{table}

\begin{tcolorbox}[colback=gray!20, colframe=gray!20, width=\columnwidth, left=0.05in, right=0.05in, top=0.05in, bottom=0.05in]
\textbf{Finding 9:}
MLLMs struggle with identifying UI design issues accurately, with an overall average accuracy of only 0.2714 across all models and frameworks.
\end{tcolorbox}

\subsection{RQ6: What mistakes do MLLMs make on the three tasks?}

We conduct systematic error analysis by randomly selecting 25\% of results for examination. The annotation process involves iterative discussion and refinement of failure categories until consensus is reached, with continuous communication to identify emerging failure types during annotation.

For the design generation task, we define seven failure categories: (1) Compile error: generated code cannot be compiled successfully; (2) Layout disorder: UI arrangement deviates from the reference design; (3) Element missing: essential UI components are absent; (4-7) Element attribute errors: wrong position, color, text content, or size of UI elements. For design edit and repair tasks, we establish six failure types: (1) Compile error: modified code contains compilation errors; (2) No edit/repair: models make no modifications; (3) Wrong objects: models target incorrect elements; (4) Wrong edit/repair: models apply inappropriate modifications to correct elements; (5) Partial edit/repair: incomplete implementation of required changes; (6) Unnecessary modification: extraneous adjustments beyond specified requirements. The detailed annotation guideline and failure type examples are in our repo\footnote{https://github.com/WebPAI/DesignBench/blob/main/assets/Failure\_Annotation\_Guideline.md}.

% We first randomly select 25\% results for analysis and then discuss, revise, and refine the failure type until everyone reaches a consensus. During annotating new data, if encountering a new failure type, annotators will communicate and update failure type in time to guide subsequent annotations.

% For design generation task, we define the following failure types: (1)
% Compile error: the code generated by MLLMs cannot be compiled correctly.(2) Layout disorder: The layout of the generated UI is inconsistent with the original layout. (3) Element missing: there are some elements missing in the generated webpage. (4-7)  Wrong element position/color/text/size: The position, color, text or size of elements in the web pages generated by MLLM are inconsistent with the original web pages.

% For design edit and design repair tasks, we define the following failure types:(1) Compile error: the code edited or repaired by MLLM cannot be compiled correctly. (2) No edit/repair: MLLMs make no edits or fixes to the UI. (3) Wrong objects: MLLMs edit or repair the wrong target element. (4) Wrong edit/repair: MLLM locates the correct element but makes an incorrect edit or repair. (5) Partial edit/repair: MLLM does not implement complete editing or repair. (6) Unnecessary modification: In addition to modifying relevant elements, MLLM makes unnecessary adjustments. 

% (7) Good edit/repair: MLLMs correctly completed the edit or repair.

\begin{figure*}[t]
    \subfigure[Design Generation.]{
    \label{fig:generation_failure}
    \centering
    \includegraphics[width = .31\textwidth]{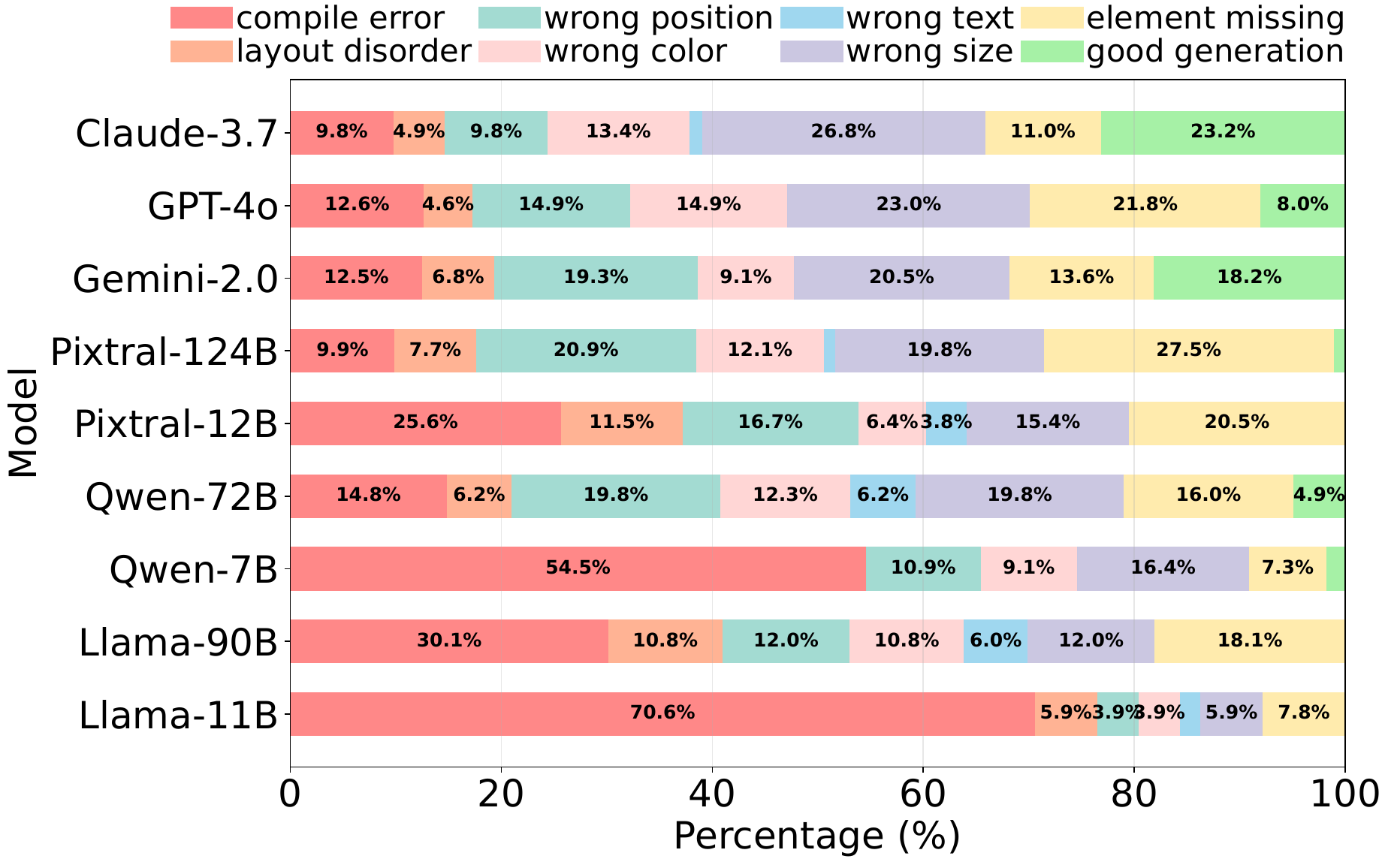}
    }
    \subfigure[Design Edit.]{
    \label{fig:edit_failure}
    \centering
    \includegraphics[width = .31\textwidth]{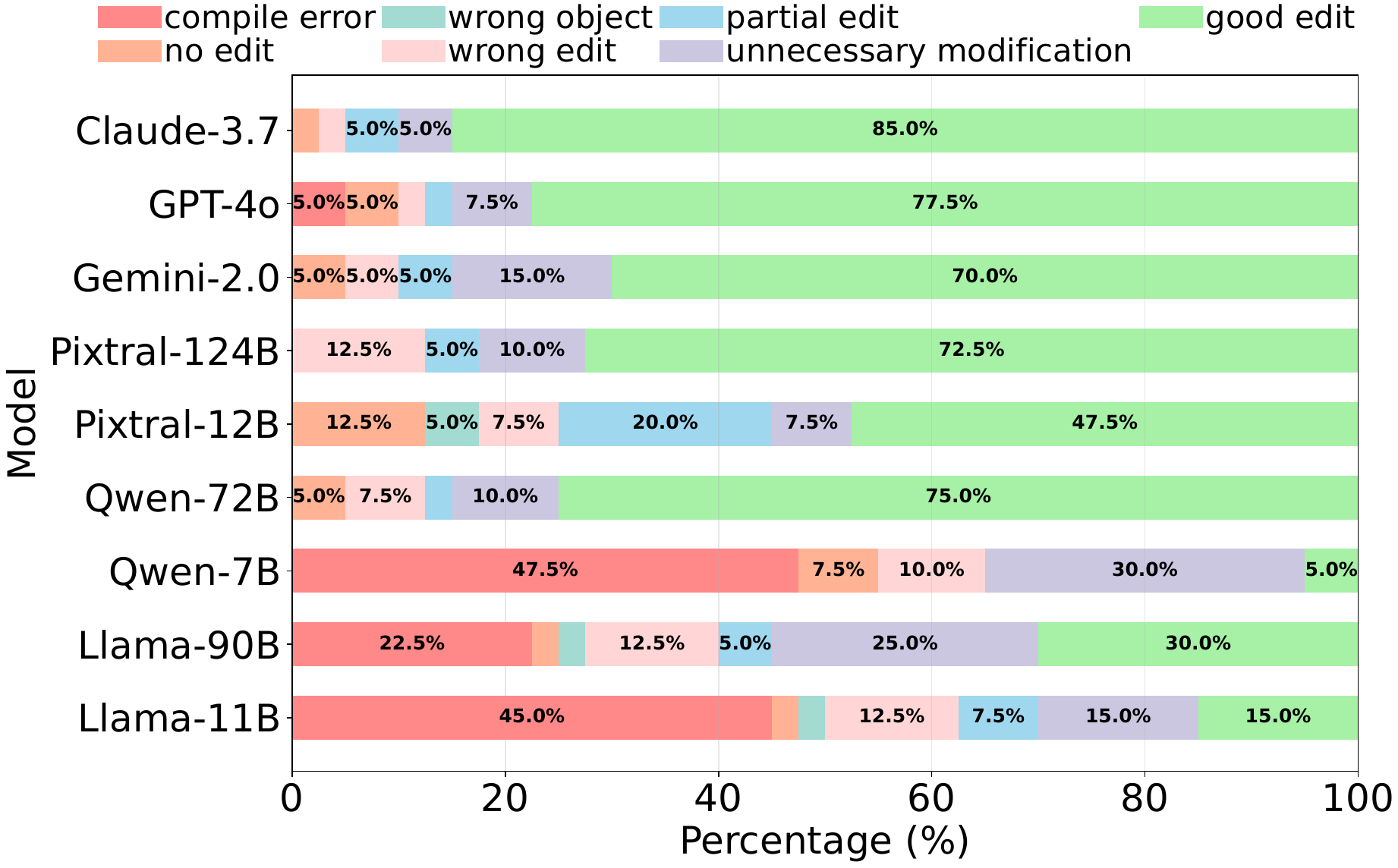}
    }
    \subfigure[Design Repair.]{
    \label{fig:repair_failure}
    \centering
    \includegraphics[width = .31\textwidth]{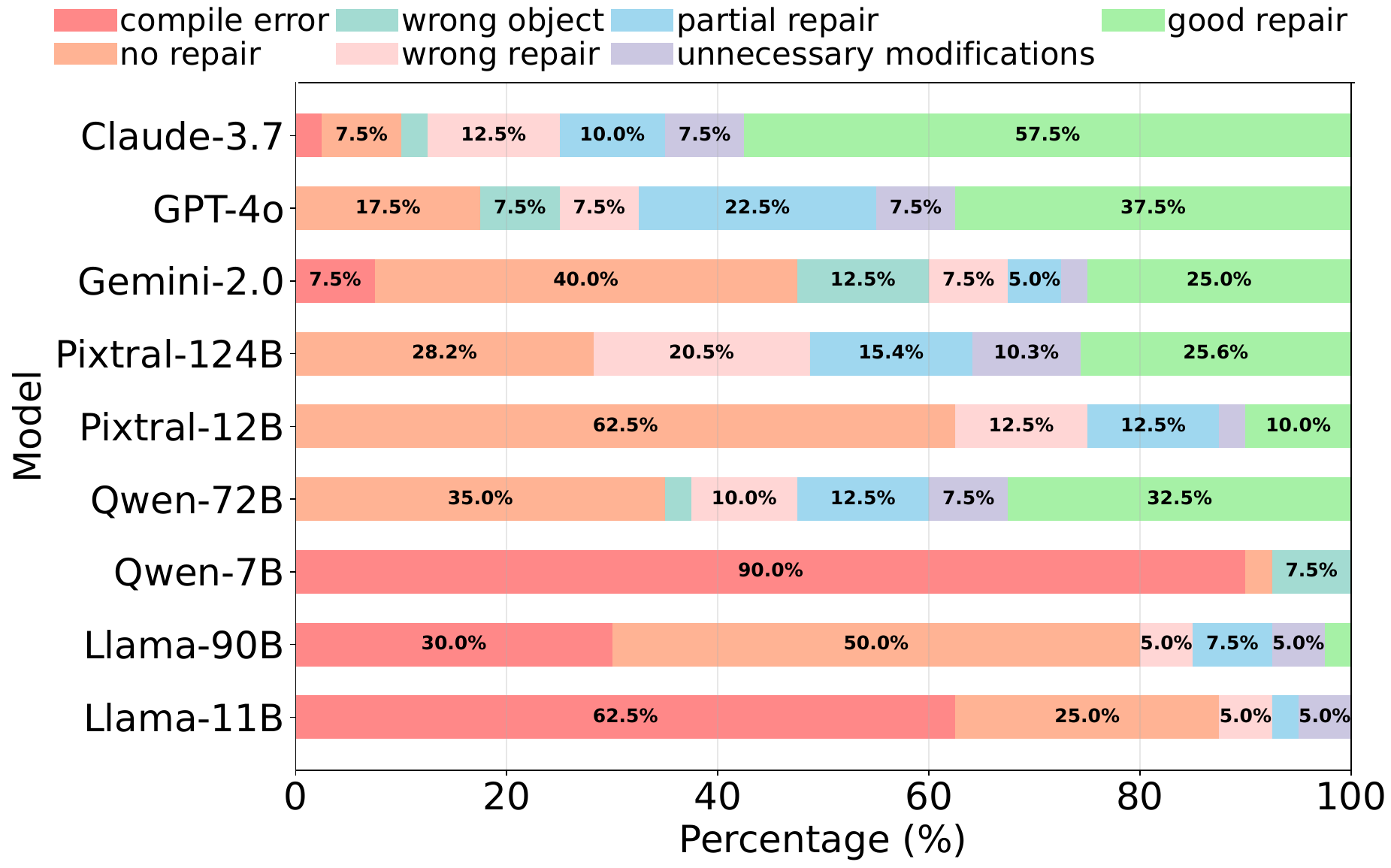}
    }
    \caption{Failure type distribution on three tasks of different MLLMs.}
    \label{fig:failures}
    % \vspace{-0.2in}
\end{figure*}

\begin{tcolorbox}[colback=gray!20, colframe=gray!20, width=\columnwidth, left=0.05in, right=0.05in, top=0.05in, bottom=0.05in]
\textbf{Finding 10:} MLLMs exhibit distinct task-specific failure patterns reflecting different cognitive demands. Design generation fails primarily on spatial reasoning (wrong element size/position) and structural completeness (element missing), design editing suffers from scope control issues (unnecessary modifications, partial edits), while design repair shows fundamental limitations in issue element identification and solution formulation (high rates of no repair attempts).
\end{tcolorbox}

From Figure~\ref{fig:failures}, we observe distinct failure patterns across tasks: (1) Design generation tasks are dominated by spatial and structural failures, with "wrong element size," "element missing," and "wrong element position" as the top three failure types, revealing MLLMs' limitations in layout comprehension and spatial reasoning. (2) Design edit tasks show more controlled failure modes, with "unnecessary modifications," "partial edits," and "wrong edits" as primary concerns, indicating challenges in precise instruction following and modification scope control. (3) Design repair tasks exhibit the most severe limitations, with "no repair," "wrong repairs," and "partial repairs" as dominant failure types, demonstrating fundamental difficulties in defect identification and corrective action formulation.

\section{Threats to Validity}

\textbf{Internal validity.} (1) The MLLM-as-a-judge scores may have reliability concerns. To eliminate this, we carefully design the prompt and provide the model with a detailed guideline of evaluation criteria. The results of human evaluation in Section~\ref{subsec:metric} also verified that the accuracy of MLLM's score is above 90\%. (2) Potential data leakage. Half of the webpages in our benchmark are sourced from closed applications (e.g., top 500 websites by Moz~\cite{top500}), and the ground truth for design repair tasks is manually written by developers, minimizing data leakage risks. Moreover, the poor framework-based code generation performance, evident in syntax errors and limited use of framework-specific features, suggests that MLLM results in the benchmark are unlikely due to mere data memorization. We also demonstrate that there is a low risk of data contamination in Section~\ref{subsec:contamination}. 

\textbf{External validity.} We only include limited frameworks of React, Vue, and Angular, due to their dominance in modern web development. These frameworks represent the majority of contemporary applications and offer diverse programming paradigms: from React's JSX to Vue's templates and Angular's TypeScript approach, providing comprehensive evaluation coverage.

% further indicates minimal data contamination effects.

% (2) Potential data leakage. A portion of our framework-based web pages are collected from closed-source applications (e.g., Mozilla websites), and the ground truth code for design repair tasks is manually written by human developers rather than sourced from publicly available repositories. This significantly reduces the likelihood that MLLMs have encountered these specific implementations during training. Furthermore, our experimental results demonstrate that models exhibit poor capabilities in framework-based code generation, evidenced by numerous syntax errors and failure to utilize framework-specific features, which further suggests that MLLMs do not memorize the code.

% Third, limited context length. As webpages become more complex with numerous interactions, the input context expands, potentially exceeding the context window constraints of MLLMs (e.g., 8192 tokens for Gemini-2.0-flash). 

% Nevertheless, this limitation can be mitigated by employing iterative generation, progressively producing interactions for a webpage over multiple rounds.

% This choice is grounded that these three frameworks are currently the most popular front-end development frameworks for developing modern website applications.Most applications in reality are developed using these three frameworks, and the diversity of webpages also poses challenges to MLLM's ability to generate web pages.
% Second, potential data leakage. 

\section{Discussion}

We elicit several actionable advice from our findings for researchers and developers in this field. \textbf{(1) For researchers:} (i) Enhance framework-specific MLLM training. The poor performance on framework-specific syntax and component-based implementations underscores the need for MLLM training datasets enriched with modern web development patterns and framework-specific best practices. This would improve their practical value for applying to diverse front-end ecosystems. (ii) Improve multimodal information fusion for UI tasks. Our findings reveal that for design edit and repair tasks, single code input and multimodal input perform equally well, suggesting that MLLMs currently underutilize visual information. Future work should prioritize developing more effective visual-code alignment for multimodal reasoning in front-end jobs.

\textbf{(2) For developers:} (i) Provide code edit location information for Design Edit and Design Repair tasks. Since MLLMs struggle with code localization, providing exact locations for edits or repairs would significantly enhance performance and reduce the cognitive burden on models to identify relevant code segments. (ii) Clearly state the repair issues for repair tasks. Given MLLMs' low accuracy in UI issue identification, explicitly stating the problems allows models to focus on targeted solutions rather than inaccurate independent diagnosis. (iii) Decompose complex instructions and large designs. Given that MLLM performance degrades with complex instructions and large UI mockups, practitioners can improve the practicality by breaking down complex requirements into simpler, atomic tasks and segmenting large UI designs into smaller, manageable components.

\section{Conclusion}

% In this paper, we propose \bench, the first multi-framework multi-task benchmark for front-end code generation. \bench \ involves three mainstream frameworks React, Vue and Angular, which are most popular for modern website development. Besides design generation task, we are the first to define the design edit and design repair task to align with real front-end development scenarios. We conduct comprehensive and extensive experiments to evaluate the performance of MLLMs during automated front-end development process from task, framework, difficulty and context dimensions. Our findings reveal the limitations of MLLM in framework-based front-end development tasks, providing clues for developers and researchers.

We introduce \bench, the first comprehensive multi-framework multi-task benchmark for front-end code generation, encompassing React, Vue, and Angular frameworks. Beyond traditional design generation, we pioneer design edit and repair tasks. Through extensive experiments across task complexity, framework compatibility, difficulty levels, contextual factors and in-depth code-level analysis, we reveal primary limitations in current MLLMs for framework-based development and elicit several actionable advice for researchers and developers.

% \section*{Data Availability}

% All the code and data are available at \url{https://anonymous.4open.science/r/DesignBench-3873}. 

% for replication and future research.

\bibliographystyle{ACM-Reference-Format}
\bibliography{arxiv}

@STRING{jan = "Jan."}

@STRING{may = "May"}

@STRING{jun = "June"}

@STRING{nov = "Nov."}

@STRING{tosem = "ACM Transactions on Software Engineering and Methodology"}

@inproceedings{
lam2025codecrash,
title={CodeCrash: Exposing {LLM} Fragility to Misleading Natural Language in Code Reasoning},
author={Man Ho Lam and Chaozheng Wang and Jen{-}tse Huang and Michael Lyu},
booktitle={The Thirty-ninth Annual Conference on Neural Information Processing Systems},
year={2025}
}

@article{gu2024cruxeval,
  title={Cruxeval: A benchmark for code reasoning, understanding and execution},
  author={Gu, Alex and Rozi{\`e}re, Baptiste and Leather, Hugh and Solar-Lezama, Armando and Synnaeve, Gabriel and Wang, Sida I},
  journal={arXiv preprint arXiv:2401.03065},
  year={2024}
}

@article{zhuo2024bigcodebench,
  title={Bigcodebench: Benchmarking code generation with diverse function calls and complex instructions},
  author={Zhuo, Terry Yue and Vu, Minh Chien and Chim, Jenny and Hu, Han and Yu, Wenhao and Widyasari, Ratnadira and Yusuf, Imam Nur Bani and Zhan, Haolan and He, Junda and Paul, Indraneil and others},
  journal={arXiv preprint arXiv:2406.15877},
  year={2024}
}

@article{jain2024livecodebench,
  title={Livecodebench: Holistic and contamination free evaluation of large language models for code},
  author={Jain, Naman and Han, King and Gu, Alex and Li, Wen-Ding and Yan, Fanjia and Zhang, Tianjun and Wang, Sida and Solar-Lezama, Armando and Sen, Koushik and Stoica, Ion},
  journal={arXiv preprint arXiv:2403.07974},
  year={2024}
}

@article{wan2024mrweb,
  title={MRWeb: An Exploration of Generating Multi-Page Resource-Aware Web Code from UI Designs},
  author={Wan, Yuxuan and Dong, Yi and Xiao, Jingyu and Huo, Yintong and Wang, Wenxuan and Lyu, Michael R},
  journal={arXiv preprint arXiv:2412.15310},
  year={2024}
}

@article{xiao2024interaction2code,
  title={Interaction2Code: How far are we from automatic interactive webpage generation?},
  author={Xiao, Jingyu and Wan, Yuxuan and Huo, Yintong and Xu, Zhiyao and Lyu, Michael R},
  journal={arXiv preprint arXiv:2411.03292},
  year={2024}
}

@article{Agile,
author = {Zhang, Sai and Xing, Zhenchang and Guo, Ronghui and Xu, Fangzhou and Chen, Lei and Zhang, Zhaoyuan and Zhang, Xiaowang and Feng, Zhiyong and Zhuang, Zhiqiang},
title = {Empowering Agile-Based Generative Software Development through Human-AI Teamwork},
year = {2025},
publisher = {Association for Computing Machinery},
address = {New York, NY, USA},
issn = {1049-331X},
url = {https://doi.org/10.1145/3702987},
doi = {10.1145/3702987},
journal = {ACM Transactions on Software Engineering and Methodology (TOSEM).},
month = jan
}

@inproceedings{wang2025can,
  title={Can LLMs Replace Human Evaluators? An Empirical Study of LLM-as-a-Judge in Software Engineering},
  author={Wang, Ruiqi and Guo, Jiyu and Gao, Cuiyun and Fan, Guodong and Chong, Chun Yong and Xia, Xin},
booktitle = { 2025 International Symposium on Software Testing and Analysis (ISSTA) },
  year={2025},
  publisher={ACM New York, Trondheim, Norway}
}

@inproceedings{designrepair,
author = {Yuan, Mingyue and Chen, Jieshan and Xing, Zhenchang and Quigley, Aaron and Luo, Yuyu and Luo, Tianqi and Mohammadi, Gelareh and Lu, Qinghua and Zhu, Liming },
booktitle = { 2025 IEEE/ACM 47th International Conference on Software Engineering (ICSE) },
title = {{ DesignRepair: Dual-Stream Design Guideline-Aware Frontend Repair with Large Language Models }},
year = {2025},
volume = {},
ISSN = {1558-1225},
pages = {646-646},
url ={https://doi.ieeecomputersociety.org/10.1109/ICSE55347.2025.00109},
publisher = {IEEE Computer Society},
address = {Los Alamitos, CA, USA},
month =May
}

@inproceedings{liu2020owl,
  title={Owl eyes: Spotting ui display issues via visual understanding},
  author={Liu, Zhe and Chen, Chunyang and Wang, Junjie and Huang, Yuekai and Hu, Jun and Wang, Qing},
  booktitle={Proceedings of the 35th IEEE/ACM international conference on automated software engineering (ASE)},
  pages={398--409},
  year={2020}
}

@article{chen2020wireframe,
  title={Wireframe-based UI design search through image autoencoder},
  author={Chen, Jieshan and Chen, Chunyang and Xing, Zhenchang and Xia, Xin and Zhu, Liming and Grundy, John and Wang, Jinshui},
  journal={ACM Transactions on Software Engineering and Methodology (TOSEM)},
  volume={29},
  number={3},
  pages={1--31},
  year={2020},
  publisher={ACM New York, NY, USA}
}

@article{liu2022nighthawk,
  title={Nighthawk: Fully automated localizing ui display issues via visual understanding},
  author={Liu and Chen, Chunyang and Wang, Junjie and Huang, Yuekai and Hu, Jun and Wang, Qing},
  journal={IEEE Transactions on Software Engineering (TSE)},
  volume={49},
  number={1},
  pages={403--418},
  year={2022},
  publisher={IEEE}
}

@inproceedings{lu2025misty,
  title={Misty: Ui prototyping through interactive conceptual blending},
  author={Lu, Yuwen and Leung, Alan and Swearngin, Amanda and Nichols, Jeffrey and Barik, Titus},
  booktitle={Proceedings of the 2025 CHI Conference on Human Factors in Computing Systems (CHI)},
  pages={1--17},
  year={2025}
}

@inproceedings{li2024mmcode,
  title={MMCode: Benchmarking Multimodal Large Language Models for Code Generation with Visually Rich Programming Problems},
  author={Li, Kaixin and Tian, Yuchen and Hu, Qisheng and Luo, Ziyang and Huang, Zhiyong and Ma, Jing},
  booktitle={Findings of the Association for Computational Linguistics (EMNLP 2024)},
  pages={736--783},
  year={2024}
}

@inproceedings{liu2025logomotion,
  title={LogoMotion: Visually-Grounded Code Synthesis for Creating and Editing Animation},
  author={Liu, Vivian and Kazi, Rubaiat Habib and Wei, Li-Yi and Fisher, Matthew and Langlois, Timothy and Walker, Seth and Chilton, Lydia},
  booktitle={Proceedings of the 2025 CHI Conference on Human Factors in Computing Systems (CHI)},
  pages={1--16},
  year={2025}
}

@inproceedings{gui2025webcode2m,
  title={Webcode2m: A real-world dataset for code generation from webpage designs},
  author={Gui, Yi and Li, Zhen and Wan, Yao and Shi, Yemin and Zhang, Hongyu and Chen, Bohua and Su, Yi and Chen, Dongping and Wu, Siyuan and Zhou, Xing and others},
  booktitle={Proceedings of the ACM on Web Conference (WWW)},
  pages={1834--1845},
  year={2025}
}

@misc{FrontendFrameworksPopularity,
  author = {Tanguy Krotoff},
  title = {Frontend Frameworks Popularity},
  year = {2024},
  howpublished = {\url{https://gist.github.com/tkrotoff/b1caa4c3a185629299ec234d2314e190}},
  note = {Accessed: 2025-05-28}
}

@misc{meta-llama,
  author = {Meta AI},
  title = {Llama-3.2-90B-Vision},
  year = {2025},
  publisher = {Hugging Face},
  howpublished = {\url{https://huggingface.co/meta-llama/Llama-3.2-90B-Vision}},
  note = {Accessed: May 2025}
}

@inproceedings{chen2024mllm,
  title={Mllm-as-a-judge: Assessing multimodal llm-as-a-judge with vision-language benchmark},
  author={Chen, Dongping and Chen, Ruoxi and Zhang, Shilin and Wang, Yaochen and Liu, Yinuo and Zhou, Huichi and Zhang, Qihui and Wan, Yao and Zhou, Pan and Sun, Lichao},
  booktitle={Forty-first International Conference on Machine Learning (ICLR)},
  year={2024}
}

@article{Qwen2.5-VL,
  title={Qwen2.5-vl technical report},
  author={Bai, Shuai and Chen, Keqin and Liu, Xuejing and Wang, Jialin and Ge, Wenbin and Song, Sibo and Dang, Kai and Wang, Peng and Wang, Shijie and Tang, Jun and others},
  journal={arXiv preprint arXiv:2502.13923},
  year={2025}
}

@article{thada2013jaccard,
  title={Comparison of jaccard, dice, cosine similarity coefficient to find best fitness value for web retrieved documents using genetic algorithm},
  author={Thada, Vikas and Jaglan, Vivek},
  journal={International Journal of Innovations in Engineering and Technology},
  volume={2},
  number={4},
  pages={202--205},
  year={2013}
}

@article{agrawal2024pixtral,
  title={Pixtral 12B},
  author={Agrawal, Pravesh and Antoniak, Szymon and Hanna, Emma Bou and Bout, Baptiste and Chaplot, Devendra and Chudnovsky, Jessica and Costa, Diogo and De Monicault, Baudouin and Garg, Saurabh and Gervet, Theophile and others},
  journal={arXiv preprint arXiv:2410.07073},
  year={2024}
}

@misc{selenium,
    title = {Selenium},
    url = {https://selenium-python.readthedocs.io/},
    author = {Selenium},
    month = {January},
    year = {2025}
}

@misc{single-file-cli,
  author = {Gildas Lormeau},
  title = {single-file-cli: Save a web page as a single HTML file},
  year = {2023},
  howpublished = {\url{https://github.com/gildas-lormeau/single-file-cli}},
  note = {Accessed: 2025-5-26}
}

@misc{DesignBench,
  author = {DesignBench Team},
  title = {DesignBench Repository},
  howpublished = {\url{https://github.com/WebPAI/DesignBench}}
}

@online{v0dev,
  title = {v0 by Vercel},
  url = {https://v0.dev/},
  urldate = {2025-05-26}
}

@online{Sketch,
  title = {Sketch},
  url = {https://www.sketch.com/},
  urldate = {2025-05-26}
}

@online{Axure,
  title = {Axure},
  url = {https://www.axure.com/},
  urldate = {2025-05-26}
}

@online{vue0dev,
  title = {Vue0},
  url = {https://www.vue0.dev/},
  urldate = {2025-05-26}
}

@online{top500,
  title = {Top 500 often visited Websites},
  url = {https://moz.com/top500},
  urldate = {2025-05-26}
}

@misc{framework,
    title = {Web Development Frameworks},
    url = {https://shakuro.com/blog/web-development-frameworks},
    author = {shakuro},
    month = {May},
    year = {2025}
}

@misc{web-framework,
    title = {Best Web Development Frameworks in 2025},
    url = {https://www.lambdatest.com/blog/best-web-development-frameworks/},
    author = {Mythili Raju},
    month = {May},
    year = {2025}
}

@inproceedings{yang2024swe,
  title={SWE-bench Multimodal: Do AI Systems Generalize to Visual Software Domains?},
  author={Yang, John and Jimenez, Carlos E and Zhang, Alex L and Lieret, Kilian and Yang, Joyce and Wu, Xindi and Press, Ori and Muennighoff, Niklas and Synnaeve, Gabriel and Narasimhan, Karthik R and others},
  booktitle={Forty-second International Conference on Machine Learning (ICLR)},
  year={2025}
}

@article{yun2024web2code,
  title={Web2Code: A Large-scale Webpage-to-Code Dataset and Evaluation Framework for Multimodal LLMs},
  author={Yun, Sukmin and Lin, Haokun and Thushara, Rusiru and Bhat, Mohammad Qazim and Wang, Yongxin and Jiang, Zutao and Deng, Mingkai and Wang, Jinhong and Tao, Tianhua and Li, Junbo and others},
  journal={arXiv preprint arXiv:2406.20098},
  year={2024}
}

@article{gui2024vision2ui,
  title={VISION2UI: A Real-World Dataset with Layout for Code Generation from UI Designs},
  author={Gui, Yi and Li, Zhen and Wan, Yao and Shi, Yemin and Zhang, Hongyu and Su, Yi and Dong, Shaoling and Zhou, Xing and Jiang, Wenbin},
  journal={arXiv preprint arXiv:2404.06369},
  year={2024}
}

@inproceedings{acsirouglu2019automatic,
  title={Automatic HTML code generation from mock-up images using machine learning techniques},
  author={A{\c{s}}{\i}ro{\u{g}}lu, Batuhan and Mete, B{\"u}{\c{s}}ta R{\"u}meysa and Y{\i}ld{\i}z, Eyy{\"u}p and Nal{\c{c}}akan, Ya{\u{g}}{\i}z and Sezen, Alper and Da{\u{g}}tekin, Mustafa and Ensari, Tolga},
  booktitle={2019 Scientific Meeting on Electrical-Electronics \& Biomedical Engineering and Computer Science (EBBT)},
  pages={1--4},
  year={2019},
  organization={Ieee}
}

@article{chen2022code,
  title={Code generation from a graphical user interface via attention-based encoder--decoder model},
  author={Chen, Wen-Yin and Podstreleny, Pavol and Cheng, Wen-Huang and Chen, Yung-Yao and Hua, Kai-Lung},
  journal={Multimedia Systems},
  volume={28},
  number={1},
  pages={121--130},
  year={2022},
  publisher={Springer}
}

@article{moran2018machine,
  title={Machine learning-based prototyping of graphical user interfaces for mobile apps},
  author={Moran, Kevin and Bernal-C{\'a}rdenas, Carlos and Curcio, Michael and Bonett, Richard and Poshyvanyk, Denys},
  journal={IEEE Transactions on Software Engineering (TSE)},
  volume={46},
  number={2},
  pages={196--221},
  year={2018},
  publisher={IEEE}
}

@article{cizotto2023web,
  title={Web pages from mockup design based on convolutional neural network and class activation mapping},
  author={Cizotto, Andr{\'e} Armstrong Janino and de Souza, Rodrigo Clemente Thom and Mariani, Viviana Cocco and dos Santos Coelho, Leandro},
  journal={Multimedia Tools and Applications},
  volume={82},
  number={25},
  pages={38771--38797},
  year={2023},
  publisher={Springer}
}

@article{jain2019sketch2code,
  title={Sketch2Code: transformation of sketches to UI in real-time using deep neural network},
  author={Jain, Vanita and Agrawal, Piyush and Banga, Subham and Kapoor, Rishabh and Gulyani, Shashwat},
  journal={arXiv preprint arXiv:1910.08930},
  year={2019}
}

@inproceedings{nguyen2015reverse,
  title={Reverse engineering mobile application user interfaces with remaui (t)},
  author={Nguyen, Tuan Anh and Csallner, Christoph},
  booktitle={2015 30th IEEE/ACM International Conference on Automated Software Engineering (ASE)},
  pages={248--259},
  year={2015},
  organization={IEEE}
}

@inproceedings{beltramelli2018pix2code,
  title={pix2code: Generating code from a graphical user interface screenshot},
  author={Beltramelli, Tony},
  booktitle={Proceedings of the ACM SIGCHI symposium on engineering interactive computing systems},
  pages={1--6},
  year={2018}
}

@misc{meta_llama,
  title = {Llama 3.2 vision},
  author = {{Meta}},
  year = {2024},
  url = {https://huggingface.co/meta-llama/Llama-3.2-90B-Vision-Instruct},
  note = {Accessed: 2025-02-06}
}

@inproceedings{papineni2002bleu,
  title={Bleu: a method for automatic evaluation of machine translation},
  author={Papineni, Kishore and Roukos, Salim and Ward, Todd and Zhu, Wei-Jing},
  booktitle={Proceedings of the 40th annual meeting of the Association for Computational Linguistics},
  pages={311--318},
  year={2002}
}

@inproceedings{radford2021learning,
  title={Learning transferable visual models from natural language supervision},
  author={Radford, Alec and Kim, Jong Wook and Hallacy, Chris and Ramesh, Aditya and Goh, Gabriel and Agarwal, Sandhini and Sastry, Girish and Askell, Amanda and Mishkin, Pamela and Clark, Jack and others},
  booktitle={International conference on machine learning},
  pages={8748--8763},
  year={2021},
  organization={PMLR}
}

@article{wang2004image,
  title={Image quality assessment: from error visibility to structural similarity},
  author={Wang, Zhou and Bovik, Alan C and Sheikh, Hamid R and Simoncelli, Eero P},
  journal={IEEE transactions on image processing},
  volume={13},
  number={4},
  pages={600--612},
  year={2004},
  publisher={IEEE}
}

@misc{openai_gpt4o,
  title = {Hello GPT-4o},
  author = {{OpenAI}},
  year = {2024},
  url = {https://openai.com/index/hello-gpt-4o/},
  note = {Accessed: 2024-10-06}
}

@misc{google_gemini_api,
  title = {Gemini API},
  author = {{Google}},
  year = {2024},
  url = {https://ai.google.dev/gemini-api},
  note = {Accessed: 2024-10-06}
}

@article{si2024design2code,
  title={Design2Code: How Far Are We From Automating Front-End Engineering?},
  author={Si, Chenglei and Zhang, Yanzhe and Yang, Zhengyuan and Liu, Ruibo and Yang, Diyi},
  journal={arXiv preprint arXiv:2403.03163},
  year={2024}
}

@article{wan2024automatically,
  title={Automatically Generating UI Code from Screenshot: A Divide-and-Conquer-Based Approach},
  author={Wan, Yuxuan and Wang, Chaozheng and Dong, Yi and Wang, Wenxuan and Li, Shuqing and Huo, Yintong and Lyu, Michael R},
  journal={Proceedings of the ACM on Software Engineering (FSE)},
  volume={1},
  number={FSE},
  pages={1--24},
  year={2025},
  publisher={ACM New York, Trondheim, Norway}
}

@article{zhou2024bridging,
  title={Bridging Design and Development with Automated Declarative UI Code Generation},
  author={Zhou, Ting and Zhao, Yanjie and Hou, Xinyi and Sun, Xiaoyu and Chen, Kai and Wang, Haoyu},
  journal={Proceedings of the ACM on Software Engineering (FSE)},
  volume={1},
  number={FSE},
  pages={1--24},
  year={2025},
  publisher={ACM New York, Trondheim, Norway}
}

@misc{laurençon2024unlocking,
      title={Unlocking the conversion of Web Screenshots into HTML Code with the WebSight Dataset}, 
      author={Hugo Laurençon and Léo Tronchon and Victor Sanh},
      year={2024},
      eprint={2403.09029},
      archivePrefix={arXiv},
      primaryClass={cs.HC}
}

@misc{anthropic_claude,
  title = {Claude},
  author = {{Anthropic}},
  year = {2024},
  url = {https://www.anthropic.com/claude},
  note = {Accessed: 2024-06-06}
}

@article{Xu2021Image2e,
  author = {Y. Xu and L. Bo and X. Sun and B. Li and J. Jiang and W. Zhou},
  title = {image2emmet: Automatic code generation from web user interface image},
  journal = {Journal of Software: Evolution and Process},
  year = {2021},
  volume = {33},
  number = {8},
  pages = {e2369}
}

@inproceedings{tang2025slidecoder,
    title = "{S}lide{C}oder: Layout-aware {RAG}-enhanced Hierarchical Slide Generation from Design",
    author = "Tang, Wenxin  and
      Xiao, Jingyu  and
      Jiang, Wenxuan  and
      Xiao, Xi  and
      Wang, Yuhang  and
      Tang, Xuxin  and
      Li, Qing  and
      Ma, Yuehe  and
      Liu, Junliang  and
      Tang, Shisong  and
      Lyu, Michael R.",
    editor = "Christodoulopoulos, Christos  and
      Chakraborty, Tanmoy  and
      Rose, Carolyn  and
      Peng, Violet",
    booktitle = "Proceedings of the 2025 Conference on Empirical Methods in Natural Language Processing (EMNLP 2025)",
    month = nov,
    year = "2025",
    address = "Suzhou, China",
    publisher = "Association for Computational Linguistics",
    url = "https://aclanthology.org/2025.emnlp-main.458/",
    doi = "10.18653/v1/2025.emnlp-main.458",
    pages = "9026--9050",
    ISBN = "979-8-89176-332-6"
}

@article{dang2025envisioning,
  title={Envisioning Future Interactive Web Development: Editing Webpage with Natural Language},
  author={Dang, Truong Hai and Xiao, Jingyu and Huo, Yintong},
  journal={arXiv preprint arXiv:2510.26516},
  year={2025}
}

@article{wan2025automatically,
  title={Automatically Generating Web Applications from Requirements Via Multi-Agent Test-Driven Development},
  author={Wan, Yuxuan and Liang, Tingshuo and Xu, Jiakai and Xiao, Jingyu and Huo, Yintong and Lyu, Michael R},
  journal={arXiv preprint arXiv:2509.25297},
  year={2025}
}

@article{xiao2025efficientuicoder,
  title={Efficientuicoder: Efficient mllm-based ui code generation via input and output token compression},
  author={Xiao, Jingyu and Zhang, Zhongyi and Wan, Yuxuan and Huo, Yintong and Liu, Yang and Lyu, Michael R},
  journal={arXiv preprint arXiv:2509.12159},
  year={2025}
}

@misc{xiao2026comuicoder,
      title={ComUICoder: Component-based Reusable UI Code Generation for Complex Websites via Semantic Segmentation and Element-wise Feedback}, 
      author={Jingyu Xiao and Jiantong Qin and Shuoqi Li and Man Ho Lam and Yuxuan Wan and Jen-tse Huang and Yintong Huo and Michael R. Lyu},
      year={2026},
      eprint={2602.19276},
      archivePrefix={arXiv},
      primaryClass={cs.SE},
      url={https://arxiv.org/abs/2602.19276}, 
}

@inproceedings{Chen2018FromUI,
  author = {C. Chen and T. Su and G. Meng and Z. Xing and Y. Liu},
  title = {From UI design image to GUI skeleton: a neural machine translator to bootstrap mobile GUI implementation},
  booktitle = {Proceedings of the 40th International Conference on Software Engineering (ICSE)},
  year = {2018},
  pages = {665--676}
}

@misc{ren2020codebleu,
      title={CodeBLEU: a Method for Automatic Evaluation of Code Synthesis}, 
      author={Shuo Ren and Daya Guo and Shuai Lu and Long Zhou and Shujie Liu and Duyu Tang and Neel Sundaresan and Ming Zhou and Ambrosio Blanco and Shuai Ma},
      year={2020},
      eprint={2009.10297},
      archivePrefix={arXiv},
      primaryClass={cs.SE},
      url={https://arxiv.org/abs/2009.10297}, 
}

% \appendix

\end{document}